\documentclass[a4paper, 11pt]{report}
\usepackage{graphicx}
\usepackage{fullpage}
\usepackage{caption}
\usepackage{subcaption}
\usepackage{lipsum}
\usepackage{tikz} 
\usepackage{float}
\usepackage{mathtools}
\usepackage{systeme}
\usepackage{siunitx}
\usepackage{glossaries}
\usepackage[absolute]{textpos}
\makeglossaries

\textblockorigin{10mm}{10mm} 
 
\usepackage[english]{babel}
\usepackage[noheader]{packages/sleek}
\usepackage{packages/sleek-title}
\usepackage{packages/sleek-theorems}
\usepackage{packages/sleek-listings}
\usepackage{stmaryrd}
\usepackage{tabularx}

\logo{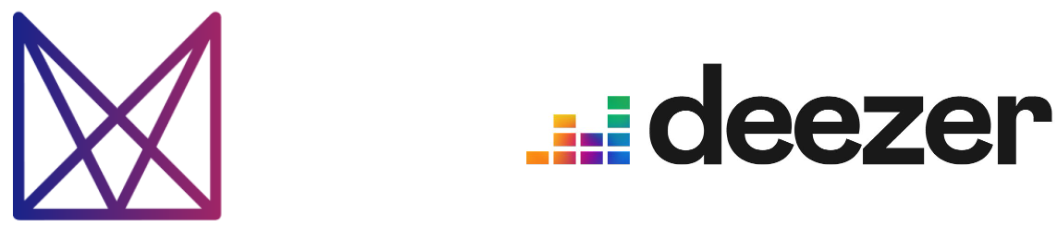}
\institute{MVA master thesis}
\faculty{Ecole normale supérieure Paris-Saclay}
\title{Audio Denoising for Robust Audio Fingerprinting}
\subtitle{Deezer Research}
\author{\textit{Author}\\Kamil \textsc{Akesbi}}
\supervisor{\textit{Tutors}  \\Benjamin \textsc{Martin} \\ Dorian \textsc{Desblancs} \\ \vspace{10pt} \textit{Referent teacher} \\ Gaël \textsc{Richard} }
\date{\today}

\lstdefinestyle{latex}{
    language=TeX,
    style=default,
    commentstyle=\ForestGreen,
    keywordstyle=\TrueBlue,
    stringstyle=\VeronicaPurple,
    emphstyle=\TrueBlue,
    emph={LaTeX, usepackage, textit, textbf, textsc}
}

\FrameTBStyle{latex}

\makeglossaries

\begin{document}
    
    \maketitle
    \newpage
    \strut
    \thispagestyle{empty}

    \vspace*{\fill}
    \begin{center}
    \begin{minipage}{.6\textwidth}
    \centering
    Kamil Akesbi \\
     \textit{Audio Denoising for Robust Audio Fingerprinting}\\
     Master MVA \\ \textit{September 2022}
    \end{minipage}
    \end{center}
    \vfill  
    \newpage

    \newpage
    \strut
    \thispagestyle{empty}
    \begin{textblock*}{8cm}(9cm,21cm) 
    \textit{Dedicated to my beloved grandmother, \\ Claude Benabdeljlil, \\
    Thank you for sharing your passion for maths and science with me} 
    \end{textblock*}
    \newpage
    
    \chapter*{Abstract}

    \hspace{0pt}
    \vfill
 
    Music discovery services let users identify songs from short mobile recordings. These solutions are often based on \textbf{Audio Fingerprinting} (AFP), and rely more specifically on the extraction of spectral peaks in order to be robust to a number of distortions \cite{Shazam}. Few works have been done to study the robustness of these algorithms to background noise captured in real environments. In particular, AFP systems still struggle when the signal to noise ratio is low, i.e when the background noise is strong \cite{Sonnleitner2017}. In this project, we tackle this problematic with Deep Learning (DL). We test a new hybrid strategy which consists of inserting a denoising DL model in front of a peak-based AFP algorithm. We simulate noisy music recordings using a realistic data augmentation pipeline, and train a DL model to denoise them. The denoising model limits the impact of background noise on the AFP system's extracted peaks, improving its robustness to noise. We further propose a novel loss function to adapt the DL model to the considered AFP system, increasing its precision in terms of retrieved spectral peaks. To the best of our knowledge, this hybrid strategy has not been tested before. 

    \vfill
    \hspace{0pt}
    
    \chapter*{Acknowledgement}

    First, I would like to express my deepest gratitude to my research supervisors, 
    Benjamin Martin and Dorian Desblancs, who gave me the opportunity to conduct this research project. Your weekly feedback regarding my propositions were very precious. You were always available, spending an important time reviewing both my report and code. This, in addition to your enthusiasm regarding this project, made me really enjoy your company.

    I would like also to thank the other members of the Deezer Research team. You passionate people were really kind to me and always available for help. You made me feel very really well integrated within the team, and I had a lot of fun with you, either during meetings, hackhatons, or at jeudrinks. 
    
    My sincere thanks also go to my professors from Ecole Normale Superieure, for sharing your passion for science with us. I have noticeably improved my research and scientific skills during this year, and this would not have been possible without the high scientific quality content and standards of your courses. I do not forget to thank my professors from Ecole Centrale de Lille, for giving me the building blocks that will allow me to become an accomplished engineer. A special thank to Mr. Pierre Chainais, who by integrating me in the \textit{SDIA} specialization, allowed me to discover the fascinating fields of machine learning and data science. 

    Finally, a big thank to my family and more precisely to my parents. None of this would have been possible without your continuous support throughout these long years of study. 

    Kamil Akesbi

    {\small\tableofcontents}

    \newacronym{AFP}{AFP}{Audio Fingerprinting}
    \newacronym{DL}{DL}{Deep Learning}
    \newacronym{MIR}{MIR}{Music Information Retrieval}
    \newacronym{RADAR}{RADAR}{Robust Algorithm for Deduplication and Audio Recognition}
    \newacronym{SE}{SE}{Speech Enhancement}
    \newacronym{ME}{ME}{Music Enhancement}
    \newacronym{stft}{STFT}{Short Term Fourrier Transform}
    \newacronym{CQT}{CQT}{Constant Q Transform}
    \newacronym{TF}{TF}{Time-Frequency}   
    \newacronym{SSIM}{SSIM}{Structural Similarity Index Measure}
    \newacronym{DFL}{DFL}{Deep Feature Loss}
    \newacronym{DRFL}{DRFL}{Deep Radar Feature Loss}
   \newacronym{TI}{TI}{Tversky Index}
   \newacronym{BN}{BN}{Background Noise}
    \printglossary[type=\acronymtype, nonumberlist, title={Acronyms}]

    {\small\listoffigures}   {\small\listoftables}

    \chapter{Introduction}
    
    \textbf{Audio identification} consists of recognizing an audio extract in a given database \cite{Sonnleitner2017}. For music streaming services, being able to identify each of the audio tracks in a unique manner, without relying on any song metadata is crucial. It allows one to quickly identify similar content and thus remove potential duplicates. At Deezer, where the music database contains more than 180 million audio tracks, this technology is used on a daily basis to help with catalog management.  
    Music identification has other applications: a popular one is the \textbf{identification of a song from a short recording}. This feature is \textbf{client oriented}: a typical scenario corresponds to a user who wants to quickly identify a song title or artist that they are listening to in a random, possibly noisy environment such as a restaurant, a car, or a shopping mall. 
    
    In recent years, different music discovery services have proposed their own solutions to let users identify songs. The most famous service is most likely \href{https://www.shazam.com/fr/home}{\textit{Shazam}}, which was launched in the early 2000s. Since then, other apps, such as \href{https://www.soundhound.com/soundhound/}{\textit{SoundHound}}, \href{https://ai.googleblog.com/2018/09/googles-next-generation-music.html}{\textit{Google Sound Search}}, 
    or Deezer's \href{https://features.deezer.com/en-us/songcatcher/}{\textit{SongCatcher}}, proposed a similar feature.  
    
    The technology behind most well-known identification apps is called  \textbf{\gls{AFP}} \cite{AudioFingerprintingConcepts}. Compact and discriminative audio features, called audio fingerprints, are extracted from a query audio segment and compared to an indexed reference database containing pre-computed audio fingerprints as well as their corresponding metadata information (title, artist). In the case of a successful identification, the content information linked to the identified fingerprints is retrieved and sent to the client. 
    
    The fingerprint extraction algorithm is deterministic, scalable and usually designed to be robust to a wide range of audio transformations \cite{RobustAFP}. Indeed, it needs to be able to identify a song from an audio recording in situations where the audio has been compressed, has undergone pitch, tempo or speed transformations, or been recorded in the presence of strong background noise.  

    Over the past two decades, different algorithms have been designed for audio fingerprinting. One popular approach concerns AFP algorithms that are based on the concept of \textbf{spectral peaks}. The AFP features correspond to characteristic points, referred to as spectral peaks, extracted from a time-frequency representation of the signal. In this work, we will refer to this type of audio fingerprints as \textbf{peak-based AFP}.
    
    In 2003, the \textit{Shazam algorithm} \cite{Shazam} appeared as one of the first successful algorithms to be scalable to databases of millions of songs, opening the way for commercial and industrial applications. Since then, improvements have been made to handle some of the audio transformations listed above. However, some cases are still challenging for current state-of-the-art algorithms. In particular, most systems still struggle when background noise is present in the recording environment \cite{Sonnleitner2017}.  
    
    Over the past few years, the emergence of deep learning (\gls{DL}) has pushed the state-of-the-art in many challenging Music Information Retrieval (\gls{MIR}) tasks such as chord and key estimation or music source separation. However, very little work has been done regarding how deep learning could be used to build noise robust fingerprinting systems. Some works (\cite{NowPlaying}, \cite{NeuralAudioFingerprint}, \cite{ContrastiveAFP}) have proposed new audio fingerprint extraction algorithms that are completely based on deep learning, but these papers propose solutions often not scalable to a catalog like Deezer's. 

    In this internship, in order to improve the identification rate in noisy environments, we test a new approach which consists of inserting a deep learning model in front of a peak-based AFP algorithm. The DL algorithm can be seen as a model which has the sole role of denoising the audio recordings, while fingerprint extraction is managed by a state-of-the-art, peaks-based AFP algorithm. To the best of our knowledge, this approach has not been tested before. 
    
    This hybrid strategy has several advantages: \textbf{1.} First, the complete pipeline (DL model + peaks-based AFP algorithm) allows us to build a system robust to noise, without loosing its ability to handle other transformations (speed, tempo, pitch, compression, ...) that are already taken into account by the AFP algorithm. \textbf{2.} Second, the DL model can be easily inserted into an AFP system already in production: neither the audio fingerprints nor the way they are stored and indexed in the reference database are modified. Only the deep learning model needs to be adapted in order to fit to the system. \textbf{3. }Finally, we keep the scalability and computation efficiency of peaks-based AFP systems. Peaks-based AFP are very light and can be indexed using fast, classic audio fingerprinting methods ( \cite{Shazam}, \cite{ReviewAFP}) . In comparison, deep learning based AFP systems often rely on features that are heavy vectorial representations, making them difficult to store, index and compare in the case of databases with million of audio tracks (\cite{NowPlaying},  \cite{NeuralAudioFingerprint}, \cite{ContrastiveAFP}). 

    In this report, more precisely, we expose the following contributions :  

    \begin{itemize}
        \item We develop a strong data augmentation pipeline that can be used to simulate realistic noise on studio recorded musics. We focus on modeling noises that are present in typical places where music is usually played. 
        \item We develop a DL denoising model that can denoise spectrogram representations of mobile phone audio recordings of music. When used in conjunction with a spectrogram peak-based fingerprinting technology, it allows to improve the AFP system robustness to noise. This proposed system, despite being developed in the context of the specific AFP developed at Deezer, is generic enough to be adapted with any peak-based AFP.
    
        \item We propose a new training strategy based on a novel loss function to adapt the denoising model more specifically to the first steps of an AFP system. This further increases the AFP system's robustness to noise. Although the loss function used is based on Deezer's specific AFP system, its definition is generic enough to be transposed to other peak-based AFP algorithm. 
    \end{itemize}

    In chapter 2, we introduce some key notions about audio fingerprinting that are essential to understand the motivations and constraints behind this work. It starts with a general presentation of audio fingerprinting systems requirements and challenges before presenting Deezer's AFP technology. In chapter 3,  we review a number of works from the audio denoising literature that inspired us to develop the DL denoising model. In chapter 4, we explore deeply this work's methodology. We present the dataset constitution, from the selection of relevant databases to the design of a strong data augmentation pipeline for noise simulation. We also outline the selected architectures for the DL denoising model, and introduce some relevant metrics used. Chapter 5 presents the trainings : we explain with which objectives the models were trained and how they were fine-tuned. In chapter 6, we expose our final results. We also discuss the limitations of our work and propose future research directions. A brief conclusion will follow.

    \chapter{Background Information}

    In this chapter, we introduce some key notions that are necessary to describe our work.  
    
    We start by conducting a review of audio fingerprinting (AFP) systems. We first introduce peak-based AFP systems : we will describe their operating principle, the properties and constraints they must meet, as well as the challenges they are still facing. We will also briefly talk about more recent systems, based on Deep Learning.

    We will then introduce Deezer's AFP algorithm, which is named \textit{Robust Algorithm for Deduplication and Audio Recognition} (RADAR) . We will present how this peak-based algorithm works and the constraints it has imposed on our denoising model.  

    \section{Audio Fingerprinting Systems}

    \subsection{General Requirements}
    
    The first audio fingerprint systems have been introduced in the early 2000s. As explained in \cite{ReviewAFP}, an audio fingerprint is a compact content based signature that summarizes an audio recording. 

    Audio fingerprinting systems are based on two fundamental processes: the \textbf{extraction}  of an audio fingerprint from an audio recording and the \textbf{search} for matches in a fingerprint database. 

    According to \cite{ReviewAFP}, extracted audio fingerprints need to fulfill multiple requirements:
    
    \begin{itemize}
        \item \textbf{Specificity}: The extracted audio fingerprints must be highly specific so that even a few seconds of audio fragment allow for discrimination over a large AFP database such as the Deezer catalog.
        
        \item \textbf{Compactness}: audio fingerprints must be small-sized so that they can be easily transmitted, compared, stored and indexed on a database. 
        
        \item \textbf{Scalability}: the computation of audio fingerprints must be simple and not excessively time-consuming in order to scale to millions of recordings. 

        \item \textbf{Robustness}: for reliable identification from mobile  recordings, audio fingerprints need to be robust to various signal distortions such as compression, equalization, pitch shifting or time scaling. They should also be robust to background noise. 
        
    \end{itemize}

    Improving a certain requirement often implies loosing performance in some other: for example, reducing the audio fingerprint size might affect the system's discrimination capabilities. The system needs to ensure a careful trade-off between the preceding requirements. 

    There are different families of audio fingerprinting algorithms. An important one concerns audio fingerprints that are based on the concept of \textbf{spectral peaks}. Spectral peaks are characteristic points in a time-frequency representation of the signal, also called spectrogram. The points are selected so that they remain unchanged in the presence of strong signal distortions. 

\begin{figure}[H]
    \centering
    \includegraphics[width=0.8\textwidth]{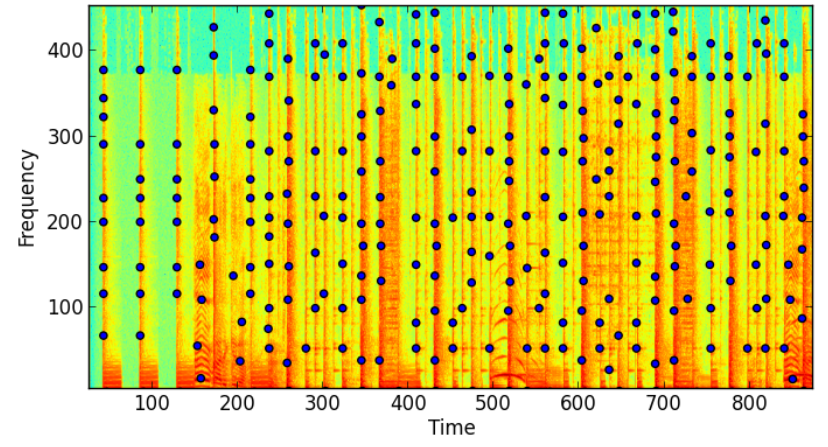}
    \noskipcaption{Spectrogram Representation of a Signal and its Spectral Peaks}
    \label{fig:constellationMap}
    \end{figure}

    Peak-based AFP were introduced by Avery Wang and constitute the basic idea behind the original \textit{Shazam algorithm} \cite{Shazam}: after computing an STFT representation of the audio signal $X$, a peak-picking strategy is used to extract time-frequency points that have the highest magnitude in a given neighborhood \ref{fig:constellationMap}. Let $\tau \in \mathbb{N}$ and $\rho \in \mathbb{N}$ determine the size of the neighborhood, a point $(n_0, k_0)$ is selected if: 

    $$ | X(n_0, k_0) | \geq | X(n,k) |$$

    for all $(n,k) \in [ n_0 - \tau, n_0 + \tau] * [k_0 - \rho, k_0 + \rho])$.

    This results in a set of coordinates that is much sparser than the original spectrogram representation. This set of coordinates is also referred to as a \textbf{constellation map} \cite{Shazam}, because of the similarity between peak patterns and  star groups in space. 

\begin{figure}[H]
    \centering
    \includegraphics[width=0.8\textwidth]{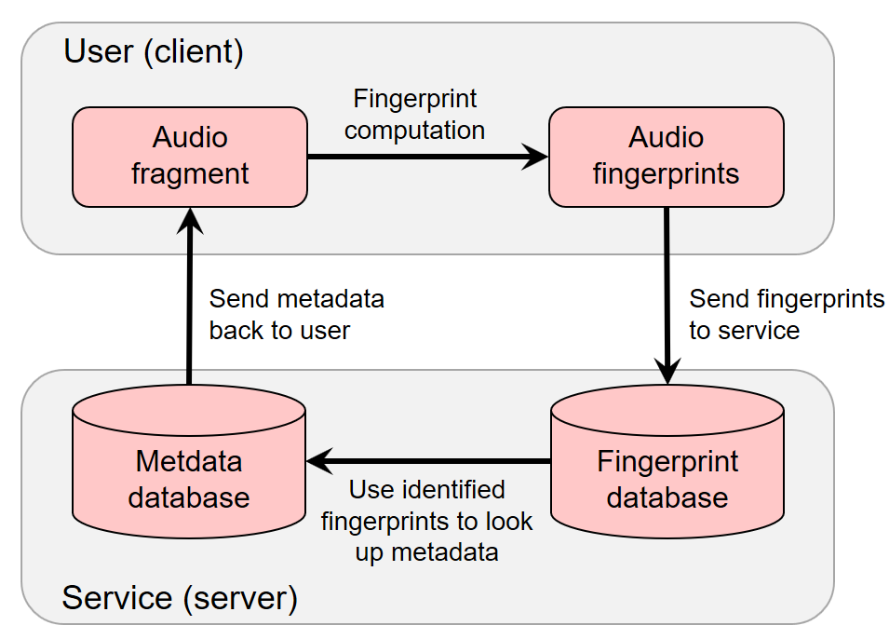}
    \noskipcaption{Audio Fingerprinting System Architecture \cite{Muller}}
    \label{fig:muller}
    \end{figure}

    For identification, audio fingerprints of the complete audio database must first be stored. The query must then be compared to all possible song sections with the same duration present in the audio database. In order to enable a fast search, the stored fingerprints must hence be indexed efficiently. The architecture of a typical AFP system is given in \ref{fig:muller}. 

    An efficient search algorithm is also needed to compare the query to the stored AFPs : we need to compare them quickly to ensure a fast identification. The indexing strategy is beyond the scope of this report, we therefore invite the reader to refer to \cite{ReviewAFP} for more details on this.

    Finally, one other key aspect to consider in audio fingerprint systems is their \textbf{granularity}, which corresponds to the number of seconds needed to identify an audio clip. The constellation must be dense enough for the matching algorithm to be able to recognize even small excerpts, but not too dense so that its indexing and storage are not too high.

    \subsection{Robustness}\label{transformations}

    Generally, audio fingerprinting systems are designed to allow the identification of pieces of music that come from the same studio mastering. Indeed, audio fingerprints based on spectral peaks are very robust to alterations of the musical piece.
    
    In particular, these systems do not generalize well to the identification of songs from live performances (a task known as cover detection). They also cannot identify songs that are not in the audio database. 

    However, in applications such as \textit{SongCatcher}, the AFP system must be able to identify the original recording of a song, even if it has undergone certain transformations. 

\begin{figure}[H]
    \centering
    \includegraphics[width=0.7\textwidth]{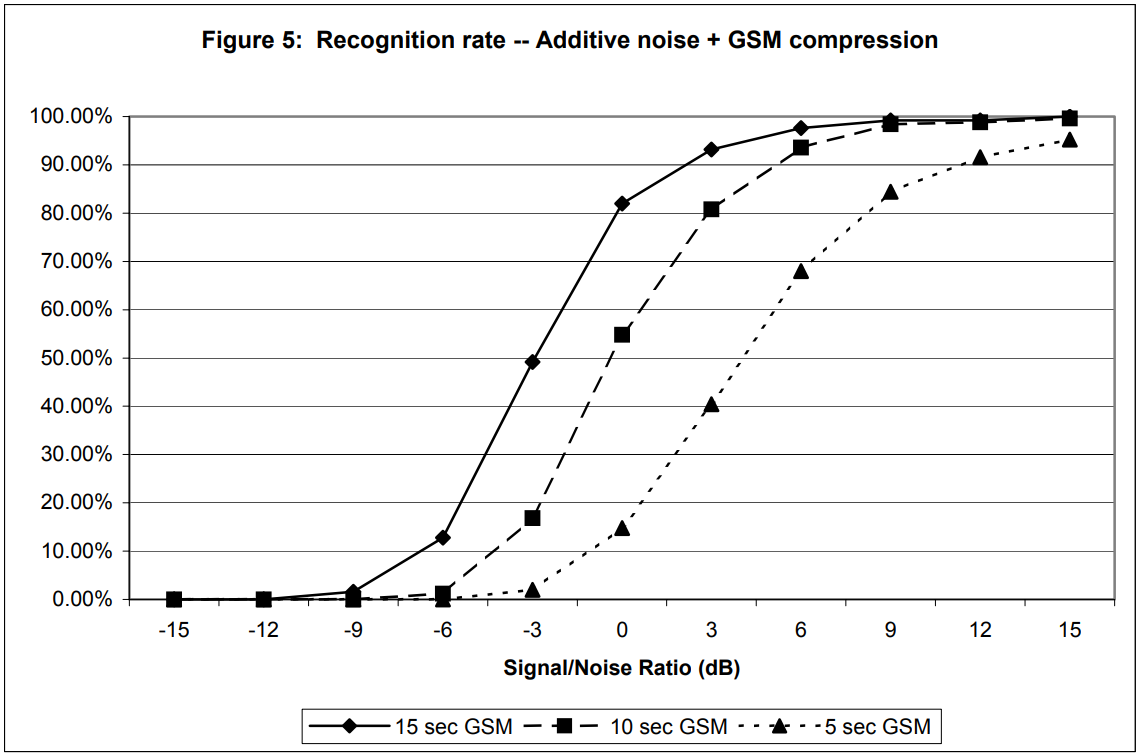}
    \noskipcaption{Shazam Algorithm Robustness to Noise}
    \label{fig:NoiseRobustnessShazam}
    \end{figure}

    In particular, the system must be robust to additive noise and distortions related to the recording device encoding. This distortions motivates the use of spectral peaks based algorithms : peaks with high amplitudes are more likely to survive in the presence of these degradations \cite{Shazam}. However, the robustness of these approaches is still limited in the presence of strong background noise, as we can see for the \textit{Shazam algorithm} in \ref{fig:NoiseRobustnessShazam}. 

    The system must also handle modifications in pitch, tempo or speed. Indeed, these transformation may be generated for instance in DJ mixes, or simply through the process of old vinyl digitization at varying playback speed.
    
\begin{figure}[H]
    \centering
    \includegraphics[width=0.6\textwidth]{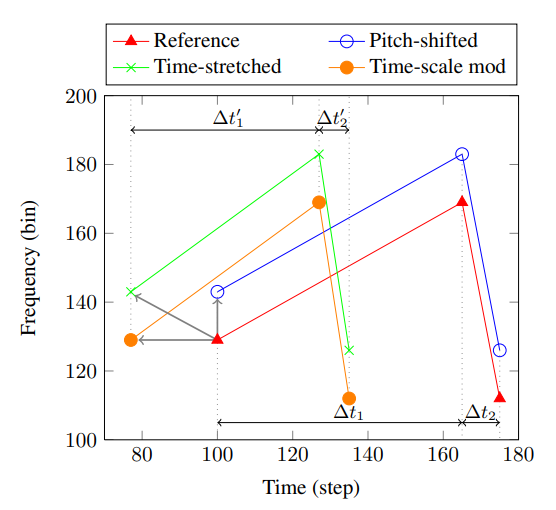}
    \noskipcaption{The Effect of Time-Scale and Pitch Modification on a Fingerprint}
    \label{fig:tripletEvent}
    \end{figure}

    Different solutions have been proposed to handle these transformations. In 2014, \cite{panako} proposed to store relative positions of triplet of peaks as fingerprints hashes to design an algorithm capable of handling time stretching, pitch shifting and time scale modifications of the signal (\ref{fig:tripletEvent}). 

    Given a triplet spectral peaks with positions $(t_1, f_1)$, $(t_2,f_2)$ and $(t_3, f_3)$, the fingerprint hash stored is: 

    $$ \Big ( f_1 - f_2, f_2 - f_3, \title{f_1}, \Tilde{f_3}, \frac{t_2 - t_1}{t_3 - t_1} \Big ) ; t_1 ; f_1 ; t_3 - t_1 $$

    Storing frequency differences and time ratios instead of absolute positions allow the system to handle the desired transformations. 

\begin{figure}[H]
    \centering
    \includegraphics[width=0.6\textwidth]{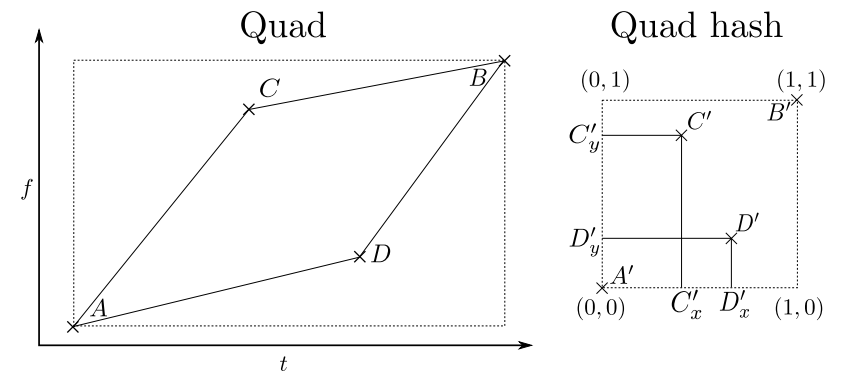}
    \noskipcaption{A Valid Quad and its Corresponding Hash }
    \label{fig:quadHash}
    \end{figure}

    One other approach, also proposed in 2014 by \cite{quad}, consists in an efficient selection and regrouping of the spectral peaks by groups of four called quads \ref{fig:quadHash}. This allows the system to handle time-scale and pitch modifications,  while increasing significantly the generated hash specificity as compared with the triplet version.

    In a typical \textit{SongCatcher} scenario, the system needs to handle recordings with strong background noise. Indeed, places where music is played (bars, restaurants, cars) often come with important non-stationary noise (people talking, moving in the area, etc.). This noise degrades the recorded content and therefore reduces the audio fingerprint system's ability to identify the correct song. 

    This type of noise is still challenging for AFP systems. To the best of our knowledge, few work has been done in specifically addressing strong real-world distortions and noise addition in the context of audio fingerprinting. 

    \subsection{Recent Approaches}

\begin{figure}[H]
    \centering
    \includegraphics[width=0.6\textwidth]{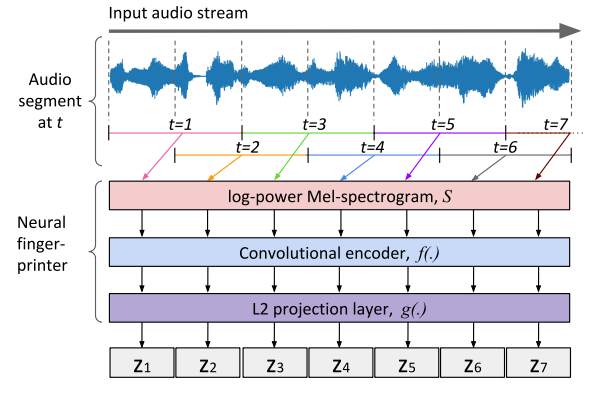}
    \noskipcaption{Deep Learning based Audio Fingerprint Extractor \cite{NeuralAudioFingerprint}}
    \label{fig:NeuralAudioFingerprinter}
    \end{figure}

    Recently, an increasing amount of audio fingerprinting works have been deep learning based. Deep learning methods have been shown to be powerful, noise-robust audio feature extractors. Features encode audio fragments and are extracted periodically, as in \ref{fig:NeuralAudioFingerprinter}.

    \emph{Now Playing by Google} \cite{NowPlaying} is one of the pioneers in this direction. This music recognizer extracts a sequence of low-dimensional vector embeddings from the audio recording using a stack of convolutional layers. In \cite{SAMAF}, the authors use a sequence-to-sequence autoencoder (SAMAF) model consisting of long short term memory (LSTM) layers to generate audio fingerprints.  

\begin{figure}[H]

    \centering
    \includegraphics[width=0.7\textwidth]
    {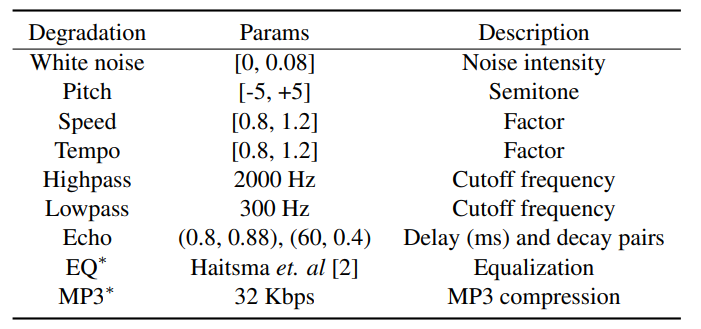}
    \caption{Augmentations used in \cite{ContrastiveAFP}}
    \label{fig:DataAugmentations}
    \end{figure}

    These algorithms are often trained using self-supervised learning strategies with contrastive or triplet losses combined with strong data augmentation pipelines \cite{ContrastiveAFP} \cite{NeuralAudioFingerprint}. The idea of these losses is to bring closer embeddings of similar samples while discriminating over dissimilar ones. In the case of audio fingerprinting, similar samples correspond to degraded versions of the same audio segment. They are generated using data augmentations corresponding to the signal distortions the AFP system must handle (see \ref{fig:DataAugmentations} for examples). 

    Although these models have been shown to be impressive compared to traditional audio fingerprinting methods, they have only been studied on relatively small datasets ($\sim$10k tracks for \textit{NowPLaying}, $\sim$100k tracks in \cite{NeuralAudioFingerprint}, $\sim$4500 hours of audio in \cite{ContrastiveAFP}). In particular, these papers do not study how the audio fingerprints are to be indexed and stored, nor do they discuss how to efficiently search the embedding space. In \cite{ContrastiveAFP} for example, the audio fingerprint of a 3 min audio track requires about 400 KB for storage. If we extend this to the complete Deezer catalog, consisting of 180 million audio tracks, it would require the system to index about 70 TB worth of data, which would be hard to sustain for fast storage indexes.

    However, there are examples of successful industrial AFP systems relying on deep learning methods. \textit{Google Sound Search}, for example, can process a musical library of over 10 million audio tracks thanks to an important spatial partitioning and vector quantization of the database and efficient search strategies. This is explained in their blog.\footnote{https://ai.googleblog.com/2018/09/googles-next-generation-music.html} These strategies require important engineering efforts which are difficult to implement and out of the scope of this internship.

    In this project, we did not develop a Deep Learning based AFP systems, but rather study how DL could be integrated in an AFP system to make the entire pipeline more robust. The literature on Deep Learning based AFP systems was still relevant to us, as it inspired us to propose realistic audio augmentation pipelines to train our denoising DL model. 
    
    \section{Introducing RADAR: Deezer's Audio Fingerprinting Technology}

    At Deezer, the audio fingerprinting system is named RADAR, for Robust Algorithm for Deduplication and Audio Recognition. It is based on spectral peak extraction and landmark construction (peaks regrouped in groups of four) to ensure robustness to various transformations.  
    We can decompose RADAR fingerprint computation into three main stages : 

    \begin{itemize}
        \item \textbf{Feature extraction}: during the first three steps, Radar applies successive filters that aim to remove most of the spectrogram information and keep only salient peaks that contain relevant information about the music track.  
        \item \textbf{Peak filtering}: during the three following steps, the algorithm filters out the extracted peaks to keep a maximum number of peaks along the time and frequency dimensions.

        \item \textbf{Landmark constitution}: peaks are regrouped into groups of four to form landmarks. This regrouping is inspired by the work done in \cite{quad}. 
    \end{itemize}

\begin{figure}[H]
    \centering
    \includegraphics[width=1\textwidth]{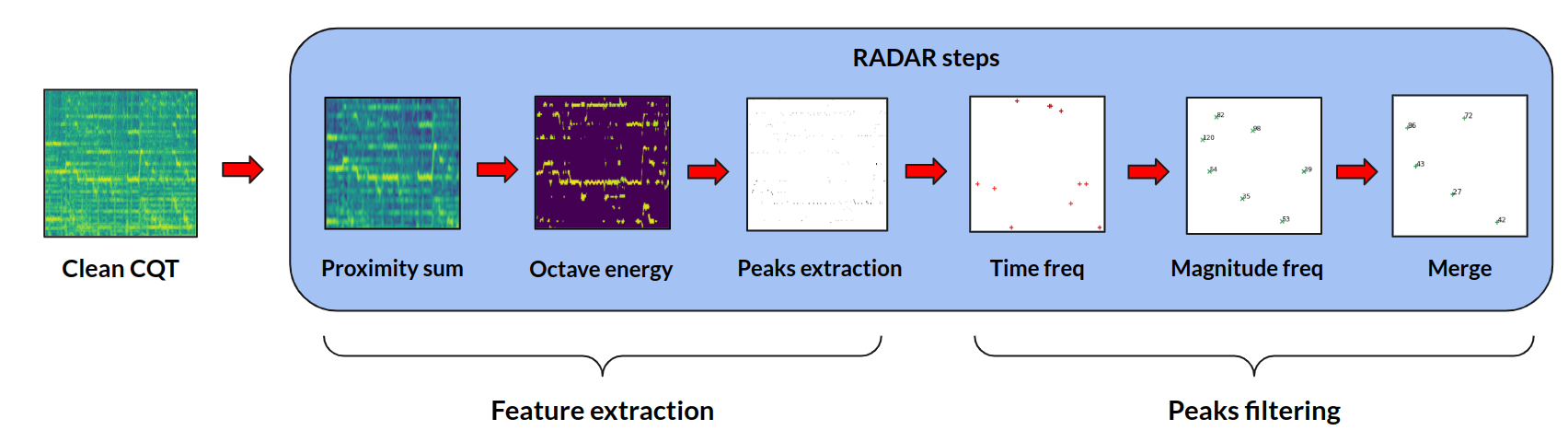}
    \noskipcaption{Radar Steps}
    \label{fig:RadarSteps}
    \end{figure}

    During this internship, we mostly focused on the first two stages of RADAR (\textbf{feature extraction} and \textbf{peak filtering}). We used them to train our model and evaluate its performance in terms of \textbf{preserved peaks} after denoising. RADAR's operations are summarised in Figure \ref{fig:RadarSteps}.

    The spectrogram processed by this algorithm is the \textbf{Constant Q-transform} (CQT). 
    The CQT provides a magnitude representation of the audio with time and frequency dimensions. The specificity of this representation is that it uses a logarithmic scale in frequencies that is well matched to the frequency distribution of music notes \cite{DL_MIR}. Center frequencies are given by:

    \begin{equation}
        f_c (k) = f_{min} * 2^{k / \beta}
    \end{equation}

    where $f_{min}$ is the minimum frequency of the spectrogram, $k$ is the integer filter index, and $\beta$ is the number of bins per octave.

    The CQT is useful for precise musical note identification. In particular, it allows AFP algorithms to handle translations in pitch, as explained in \cite{CQT-AFP}. However, one drawback of the CQT is that it is heavier to compute than other commonly used spectrograms such as mel-spectrograms. It is also more difficult to invert. 

    RADAR can handle most of the signal distortions mentioned in \ref{transformations} : the system can handle modifications in tempo, pitch and speed. It is also robust to MP3 compression and remastering modifications such as equalization. 

    Nevertheless, the algorithm is still sensitive to noise, as we can see in \ref{fig:noise}. Even though Radar is more robust then other open source AFP models (for example, \textit{chromaprint} in Figure \ref{fig:noise}), its identification performance decreases quite rapidly in the presence of brown noise : 

    \begin{figure}[H]
    \centering
    \includegraphics[width=1\textwidth]{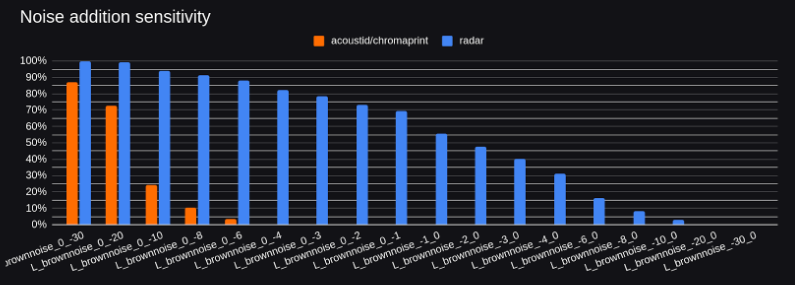}
    \noskipcaption{Robustness to Brown Noise: RADAR vs Chromaprint}
    \label{fig:noise}
    \end{figure}

    In this project, we do not aim to develop a new AFP system from scratch, but rather adapt and integrate a DL model in the AFP algorithm pipeline. This allows the system to be more robust to noise while retaining all its other benefits, such as indexation and search efficiency.
    
    \chapter{Related work}

    In this section, we present the literature that inspired us to develop a DL denoising model. During this internship, Literature review has been centered around our main objective, which is the developement of a music denoising model, keeping in mind that our model comes with certain constraints : 

    \begin{itemize}
        \item First, the model needs to be quite \textbf{light} and have a \textbf{low inference time}. Indeed, it should be inserted in an AFP system (potentially integrated on a mobile device app) which will need to perform live identification in just a few seconds. The DL model should therefore not slow down too much the system. 

        \item Moreover, we need to keep in mind that RADAR is based on the Constant Q-transform spectrogram, which might restrict the input data the model processes. 

        \item Finally, one important aspect to keep in mind is that we are dealing with music, which is an audio type that has specific properties that might differ or be more complex then other audios types often studied in the literature such as speech. In particular, we observed that the audio denoising literature focuses mostly on speech and there are very few papers tackling the topic of music denoising.  

    \end{itemize}

    \section{Audio denoising}

    \subsection{Common architectures} \label{SE}

    During its acquisition, an audio signal can be corrupted by different types of noise. Audio denoising, or enhancement, aims to attenuate the noise present in the recorded audio without modifying the original signal. 
    
    This topic has been studied for decades. Traditional approaches for audio enhancement were based on signal processing techniques such as spectral substraction \cite{SpecralSubstraction}, Wiener Filtering \cite{WienerFiltering} or also bayesian estimators \cite{BayesianEstimator}. These methods have been principally tested on speech audios and often require a prior estimation of the signal to noise ratio based on an initial silent period of the signal. As a consequence, they have trouble generalizing to non-stationary noises such as real background noise.  

    Recently, Deep Learning-based methods appeared in \textbf{speech enhancement} and have shown significant improvements over traditional methods. These approaches are data driven : clean-noisy audios pairs are used to train a denoising model \cite{DenoisingAE}. The model takes as input a noisy representation of the audio and learns to return its clean version. These approaches are supervised, since the model seeks to minimize a reconstruction error between an estimated audio representation and its clean reference. They have shown better generalization capabilities to non-stationary noises and generate higher speech quality. 

    \begin{figure}[H]
    \centering
    \includegraphics[width=0.8\textwidth]{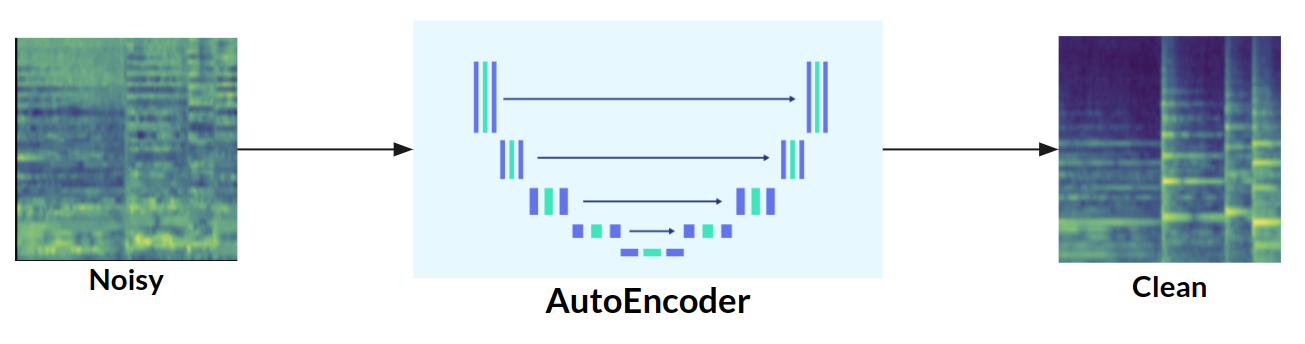}
    \noskipcaption{Training an Autoencoder with Clean-Noisy Spectrogram Pairs}
    \label{fig:AE}
    \end{figure}    

    The Deep Learning architectures used in speech Enhancement are \textbf{Auto-encoders}. These models learn how to efficiently compress and encode an input data into a small dimensional representation and how to bring it back to its original dimension (\ref{fig:AE}). Depending on the training paradigm, these models can be used to learn meaningful data representations or to perform reconstruction and denoising. Over the years, different autoencoder architectures have been used and adapted to speech enhancement : in 2013, the \textbf{deep denoising auto-encoder} was trained on clean-noisy speech pairs \cite{DenoisingAE}, \cite{DenoisingAE2}. Long short term memory (LSTM) have also been used to incorporate temporal structures in the denoising process \cite{LSTM}.
    
\begin{figure}[H]
    \centering
    \includegraphics[width=0.5\textwidth]{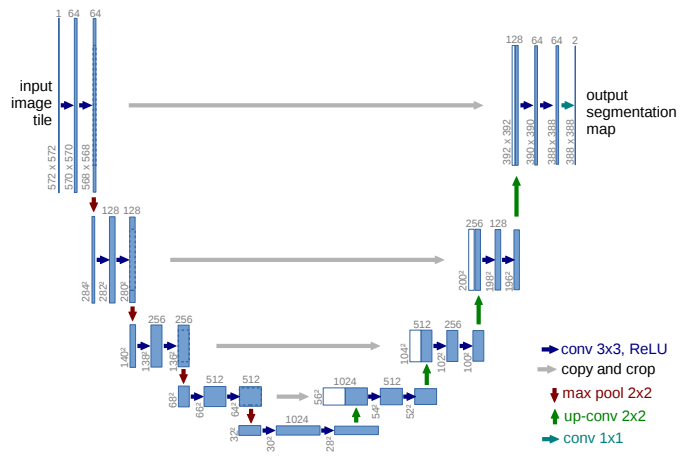}
    \noskipcaption{The original U-net architecture}
    \label{fig:U-net}
    \end{figure}

    The \textbf{U-net} architecture (\ref{fig:U-net}) finally appeared as a powerful architecture to perform speech Enhancement : this deep autoencoder model can learn to denoise the input data while keeping fine grained details about the input data thanks to its skip connections. Skip connections in U-nets, as the name suggests, skip layers from the encoder in the neural network and feed directly the corresponding layers with same dimensions in the decoder (\ref{fig:U-net}). This architecture has been first introduced in 2015 in the field of Computer Vision to perform medical image segmentation \cite{U-net}. 

    This model is fully convolutional, allowing it to take images of varying shapes as input. In the encoder part, 3*3 CNN layers are applied, followed by ReLu activations. During downsampling steps, 2*2 max pooling operations with stride 2 are used and the number of feature channels is doubled. In the decoder, feature maps are upsampled (using an 2*2 \textit{UpConvolution} layer or a \textit{ConvTranspose2D} layer) and the number of feature channels is divided by two. The upsampled outputs are concatenated with the feature maps from the encoder and then processed by a 3*3 convolution followed by a ReLu activation. In the final layer, a 1*1 convolution is finally used to map the last layer features to a single image with same dimensions as the input data.  

    EncoderDecoder models with skip connections (U-net like models) have been widely studied and adapted in the fields of audio denoising and enhancement : \cite{WaveUnet}, \cite{Demucs}, \cite{Segan}, \cite{AeGAN}, \cite{LearningToDenoiseHistoricalMusic}, \cite{TwoStageUnet} obtaining often state of the art results. The model have also received a strong attention in the field of \textbf{music source separation} (\cite{Spleeter}, \cite{DemucsMusicSeparation}, \cite{UnetMusicSourceSeparation}), indicating that it is powerful enough to understand complex audios structures such as music. 

    In the last years, there has been an increasing interest over attention based models such as the transformer in the Machine Learning community. This is also observed in the speech enhancement literature \cite{SepFormer}, \cite{SETransformer}, \cite{TSTTN}. More recently, \textbf{dual-path transformers} as well as \textbf{conformers} have shown very interesting increases in performance, achieving state of the art results in 2022 \cite{CMGAN}, \cite{DPT-FSNet}. These models combine different deep learning blocks (convolution and attention) to efficiently capture both local and global features of the input sequences. 

    \subsection{Speech Enhancement}
    
    Speech Enhancement (\gls{SE}) literature constitute a major portion of the audio denoising literature spectrum. It aims to improve the quality of noisy speech.  SE techniques can be categorized into time and time-frequency (\gls{TF}) domain methods. 
    
    \begin{itemize}
        \item \textbf{Time frequency domain models} process a spectrogram representation of the speech audio : they learn to denoise spectrograms, which can then be inverted to get the corresponding waveform signal. They constitute the classic SE paradigm : most SE models have been developed over this framework \cite{DenoisingAE}, \cite{DenoisingAE2}, \cite{CMGAN}. Historically, these models were only trained to process the magnitude component of a spectrogram, ignoring the signal's phase. This is due to the the random and complex structure of the phase, which impose challenges to DL architectures. Some works have still proposed to incorporate the noisy phase with the denoised spectrogram for waveform reconstruction \ref{fig:phase} : 

        \begin{figure}[H]
        \centering
        \includegraphics[width=0.8\textwidth]{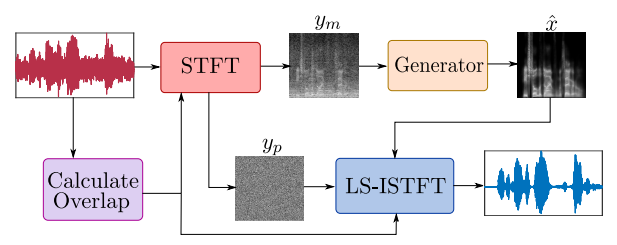}
        \noskipcaption{AeGAN \cite{AeGAN} : a Magnitude Denoising Model Incorporating the Phase }
        \label{fig:phase}
        \end{figure}

        Typical spectrograms used in speech enhancement are the Short Term Fourier Transform (STFT) ( as in \cite{CMGAN} \cite{AIAT}) or the mel-spectrogram (in \cite{melGAN}, \cite{MusicEnhancement}). The constant Q-transform, being adapted to music processing, is absent from the speech enhancement literature. More generally, we noticed that the \gls{CQT} is not used in the audio denoising literature, even for music applications.  

        More recently (since 2021), some approaches followed the strategy of enhancing the \textbf{complex spectrogram}, implicitly relying on both magnitude and phase information to perform denoising (\cite{CMGAN}, \cite{AIAT}). These methods currently match state of the art performance in speech enhancement (2022) (see \ref{fig:noise}). One drawback of these models is that they are often quite complex and heavy, making them difficult to use for real time processing. 

        \item \textbf{Time domain models} directly learn to estimate samples of the clean audio waveform from their noisy counterparts. In this case, the complete audio information is preserved in the waveform and processed by the model. However, the lack of frequency representations can make it harder for the model to capture speech phonetics in the case of SE, or music representations in the case of \gls{ME}. Indeed, as explained in \cite{MusicEnhancement}, it is probably easier to denoise music in the TF domain as it is polyphonic : additive sources (corresponding to vocals and instruments) usually cover different regions of the frequency spectrum and can be thus more easily identified using the frequency features of the signal.  Among the most important waveform models, one can quote \textit{WaveNet}  \cite{WaveNet} (2018), which uses dilated convolutions with exponentially increasing receptive field to process the raw waveforms, the \textit{Demucs} architecture (2020), which can run in real time on a laptop CPU, or more recently \textit{MANNER} \cite{MANNER} (2022), which shows state of the art results among time domain speech enhancement models.   
    \end{itemize}  

    \begin{figure}[H]
    \centering
    \includegraphics[width=0.8\textwidth]{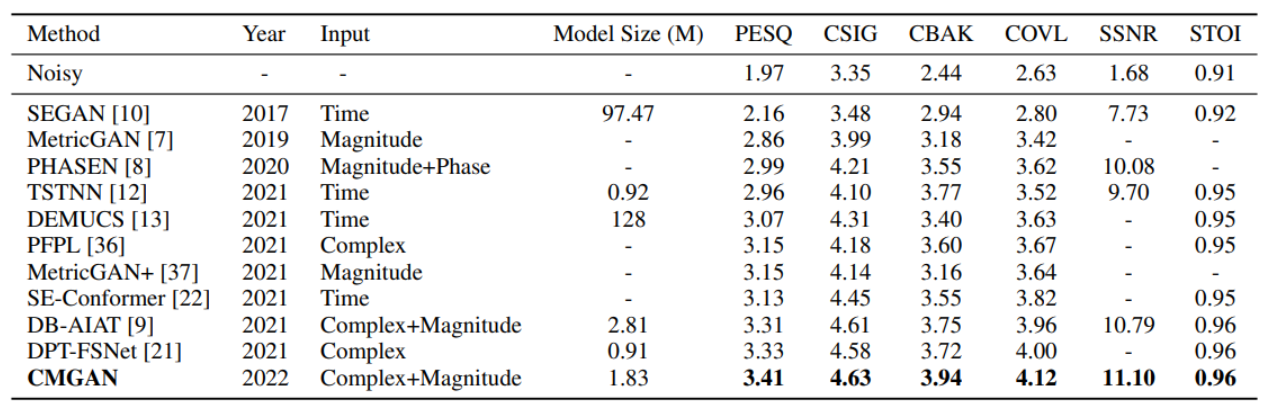}
    \noskipcaption{SOTA SE Model Performance Comparison on Voice Bank + DEMAND datasets }
    \label{fig:noise}
    \end{figure}

    \subsection{Common losses in audio denoising}\label{losses}
    
    Different losses are used in denoising. In this work, because the developed models operate on the T-F domain, both audio denoising and image restoration literature losses have been considered :  

    \begin{itemize}
        \item $L_p$ losses have been widely used in  audio denoising and image restoration. The most used one are the \textbf{mean absolute error} (L1 - MAE) loss and the \textbf{mean square error} (L2 - MSE). In speech enhancement, these losses have been used in both time and time-frequency domain. In the TF domain, they are often referred to as \textbf{spectral losses}. They are also often used in music source separation (\cite{Spleeter}, \cite{DemucsMusicSeparation}). They constitute the most simple losses for denoising and can reach great performances. However, these losses don't necessarily reflect the natural biases in human hearing. In particular, they fail to correlate with the human judgement of speech quality and intelligibility \cite{MetricGAN}. 
        
        \item \textbf{Perceptual losses} are another kind of losses used in both audio and image denoising. The specificity of these losses is that they compare high level image features instead of only comparing images pixel by pixel. The \textbf{\gls{SSIM} loss}, from the image processing literature, attempts to quantify the visibility of errors between a distorted image and its clean reference using a variety of known properties of the human visual system \cite{SSIMLoss}. Another important loss is the \textbf{Deep Feature loss} \cite{DeepFeatureLoss} : it uses a pretrained Deep Learning with fixed weights model as a feature extractor. The goal is to make the features of the prediction closer to the features of its ground truth. The feature extractor can be used to propagate gradients to the model being trained. This strategy has shown great results in SE and image denoising : the features learned by Deep Learning models are generally good at capturing  meaningful and discriminative informations, making them useful for measuring perceptual similarity.  \textit{Predicted} and \textit{ground truth} features are usually compared using  classic $L_p$ losses.  In image processing, one common DL model used as \gls{DFL} is the VGG model trained on image-net dataset. 
        \item Finally, \textbf{Adversarial losses} are also often used in \textbf{speech enhancement}, allowing for significant gains in performance. During the last years, it appeared as the most used framework in speech enhancement (\cite{Segan}, \cite{MetricGAN}, \cite{MetricGan+}, \cite{AeGAN}, \cite{HifiGan}, \cite{CMGAN}). Indeed, adversial trainings allow models to generate outputs that look "real". This is particulary interesting in the case of audio and image denoising to remove potential artifacts from datas generated by models trained with more traditional losses. One drawback of this loss is that it can be quite unstable, making it difficult to train.         
    \end{itemize}

    \section{Denoising music : an understudied problem} \label{MusicDenoising}

    Music denoising is not as studied as speech enhancement. 
    
    Some previous works have studied how to denoise \textbf{historical recordings} (\cite{TwoStageUnetHistRec}, \cite{LearningToDenoiseHistoricalMusic}) :  indeed, due to technological constraints, early recordings can have poor quality and be affected by alterations such as hiss and clicks \cite{AudioRestoration}. This research field therefore aims to restore old analog discs. Recent approaches, based on DL, also rely on U-net architectures (\cite{TwoStageUnetHistRec}, \cite{LearningToDenoiseHistoricalMusic}), as in SE or music source separation. This shows, once again, that U-net architectures are great encoder-decoders to learn from music data. 

    A recent publication \cite{MusicEnhancement} (2022), appeared as the first paper to introduce the notion of \textbf{music enhancement} (ME) : in their work, the authors studied how low-quality music recordings, such as the ones captured by mobile devices, could be enhanced using DL. This work was very inspiring to us, since our objectives were quite similar. In particular, we were interested by the data degradation pipeline and the DL architectures developed. 
    
    However, the final goal of each work is different : in music enhancement, the goal is to reconstruct denoised audios after denoising mel-spectrograms. For this purpose, the authors developed a music vocoder model for mapping synthetically generated mel-spectrograms to realistic waveforms in addition to a time frequency denoising model \ref{fig:ME} : 

    \begin{figure}[H]
    \centering
    \captionsetup{justification=centering}

    \includegraphics[width=1\textwidth]{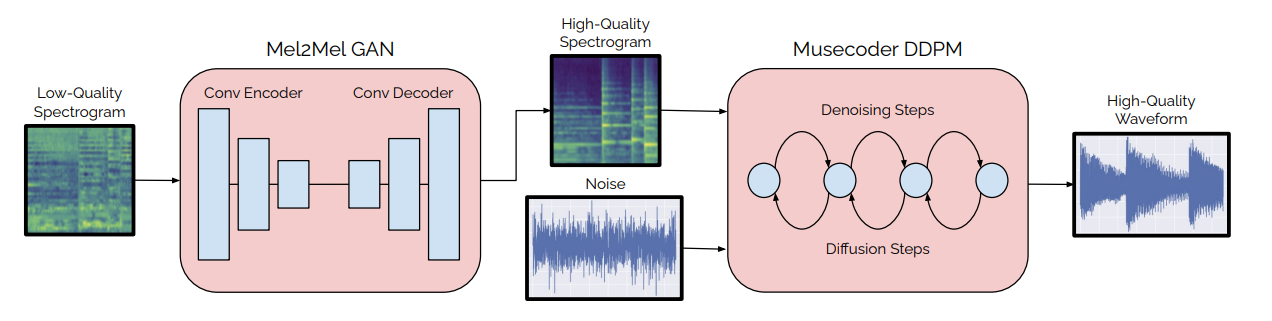}
    \noskipcaption{Music Enhancement Architecture : Denoising (Mel2Mel GAN) + Vocoder (Musecoder)}
    \label{fig:ME}
    \end{figure}

    In our case, we do not seek to reconstruct the audios waveforms : our primary objective is to denoise the CQTs of the recorded audios in order to preserve as many spectral peaks as possible.   

    \chapter{Methodology}

    In this chapter, we present the methodology and main contributions of our work. 
    
    \section{Noise simulation}

    \subsection{Sources of noise identification}

    A first important part of our work lies in the careful design of an audio augmentation pipeline that could simulate the different signal distortions we want the denoising model to handle. This pipeline was used to apply distortions to the clean audio tracks of the selected audio database. 

    We had to carefully choose and define the transformations to integrate in the augmentation pipeline. As described before, the scope of our work is restricted to removing real-world alterations. Contrary to \cite{NeuralAudioFingerprint} or \cite{ContrastiveAFP} (see \ref{fig:DataAugmentations}), we choose not to include pitch, speed or tempo transformations, as they are already handled by the system we want to improve. 

    Here, the DL model must only serve to remove noise and recording artefacts from the recordings while preserving music. As a consequence, the augmentation pipeline must only serve to simulate noise that is usually recorded in typical places where \textit{SongCatcher} is used. The first question we tried to answer was therefore : \textbf{what are the different sources of noise in typical places where \textit{SongCatcher} is used ?} 

    To answer this question, we first found inspirations in Deep Learning based audio fingerprinting systems papers, where strong data augmentations pipeline are applied to the training datas
    (\cite{NeuralAudioFingerprint}, \cite{ContrastiveAFP}). We also relied on the thesis \textit{"data augmentation techniques for robust audio analysis"} \cite{DataAugmentationThesis} to find how noise could be modeled in real environments. 
    
    In the case of recordings obtained with \textit{SongCatcher}, we identified four sources of noise:

    \begin{itemize}
        \item \textbf{Room noise} - The sound waves played in a room reverberates on its walls and shapes, creating audio artefacts in the recordings which can be more or less important depending on the room's geometry. Reverberation can be simulated by doing the \textit{convolution} of the room impulse response with the signal. 
        
        \item \textbf{Background noise} - it corresponds to additive noise summed to the original signal. It can be Gaussian noise (stationary noise) or noise from background recordings, such as acoustic scene samples. Acoustic scene recordings contain non-stationnary events, which are expected to appear in real noisy datas.  

        \item \textbf{Recording device noise} - the recording device can also distort the signal : frequency responses may vary from a mobile phone to another, cutting some frequency regions of the recording. Recorded audios can also be resampled and compressed by the recording device. They may also be subject to various gains and artifacts such as clipping. 
        
        \item \textbf{Speakers} - Loud speakers have different frequency responses. Convolving the original music recording with the impulse response of the speakers can simulate loud speaker "degraded" output. 
    \end{itemize}

    The first three sources are common in speech and audio denoising literature, to model for example the surrounding noise of people talking. In the case of music, the AFP of tracks stored correspond to features extracted from high quality original recordings. Any decline in quality of the recording can be seen as noise. In particular, loud speakers playing music in bars, cafes or cars can alter the music quality and must be thus considered as a source of noise. This is why we added \textbf{speakers} as a source of noise in our pipeline. 

    This summarizes the final augmentation pipeline, in which each source of noise affects the audio in a specific order given by \ref{fig:NoiseSoures} : 

    \begin{figure}[H]
    \centering
    \includegraphics[width=1.05\textwidth]{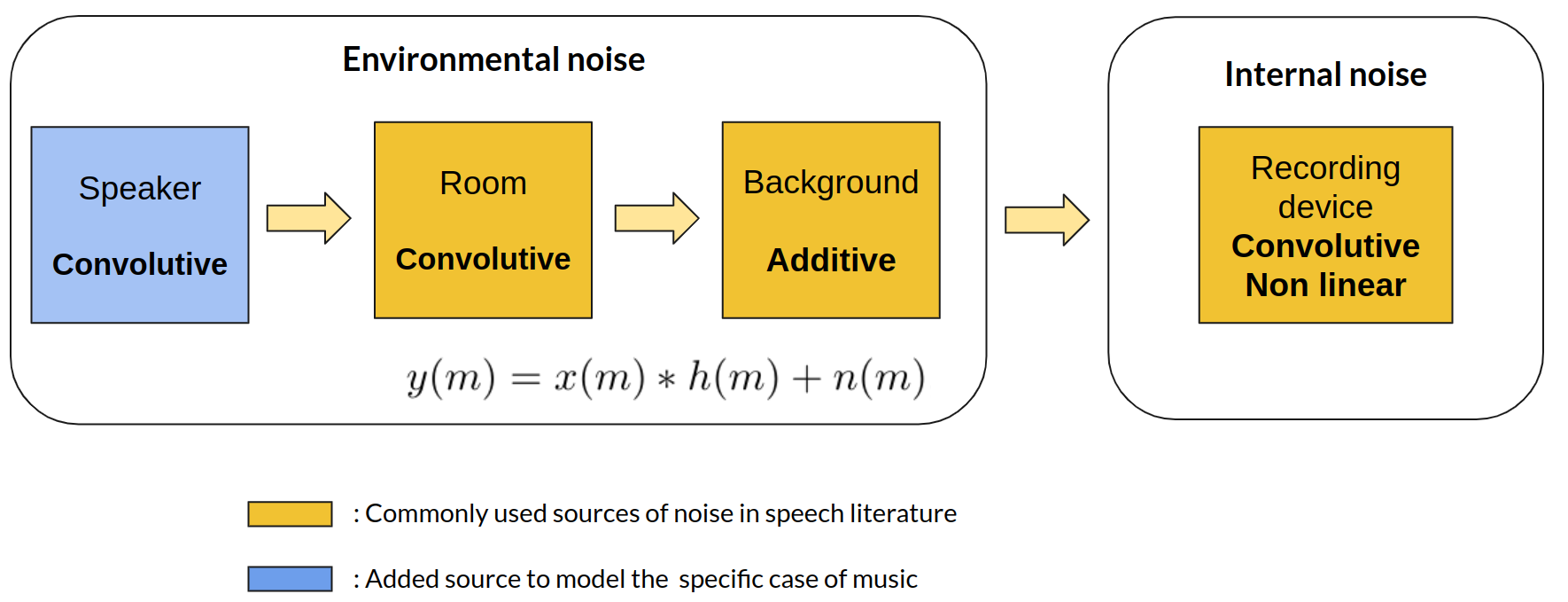}
    \noskipcaption{Selected Sources of Noise for Music Denoising}
    \label{fig:NoiseSoures}
    \end{figure}

    The speaker, Room and Background noise correspond to environmental noise. They do not depend on the recording device. Room and background noise can be modeled by the following formula : 

    $$ y(m) = x(m) * h(m) + n(m) $$

    where $y(m)$ is the distorted signal, $x(m)$ is the clean signal, $h(m)$ is the room impulse response and $n(m)$ the additive background noise.

    \clearpage
    \subsection{Noise datasets selection}
    
    After source identification, we selected relevant \textbf{noise datasets} to modify the clean tracks with realistic noise.  
    
    \subsubsection{Selecting a background noise dataset: } \label{BN}

    It is important to choose a background noise dataset which contains typical sounds that might be heard in places where \textit{SongCatcher} is used. \textit{SongCatcher} is mostly used in places where music is played, as for example : 

    $$ \textbf{Café - Restaurant - Night club - Car (radio) - Street - Malls}$$

    These types of recordings can be found in \textbf{acoustic scene datasets}. These datasets generally contain audio of different lengths, recorded in different places and cities, and are great for modeling non stationary additive noise. \footnote{An important list of such datasets can be found here : \url{https://dcase-repo.github.io/dcase_datalist/datasets_scenes.html}}. 

    We identified two sources of relevant datasets for our task : 

    \begin{itemize}
        \item \textbf{WHAM!} : Contains about 70 hours of background noise recorded at various urban locations  throughout the San Francisco Bay Area \cite{WHAM}. This unlabeled dataset was of interest to us because the recordings are mostly taken from restaurants, cafes, bars, and parks. However, after careful checks on some of the audios, we noticed that background music was present on many samples, making the data set unusable for additive noise simulation on clean music tracks. 

        \item \textbf{DCASE challenge} : The Detection and Classification of Acoustic Scenes and Events (DCASE) challenge is an annual competition proposing different tasks such as Acoustic scene classification. It comes each year with a new or expanded dataset of acoustic scenes, recorded in different places, cities and with different devices (\cite{DCASE_challenge_2018}, \cite{DCASE_challenge_2020}). 
    \end{itemize}

We thus selected DCASE datasets from the 2017 (\textit{TUT Acoustic scenes 2017 dataset}), 2018 (\textit{TUT Urban Acoustic Scenes 2018 Mobile dataset}) and 2020 (\textit{TAU Urban Acoustic Scenes 2020 Mobile dataset}) editions. Each edition comes with a \textit{development} and an \textit{evaluation} dataset, containing  \textbf{10 seconds} segment recordings.

The 2018 and 2020 datasets contain less acoustic scene classes than the 2017 edition, but have a much larger number of audio samples, recorded in six different cities accross Europe,  adding variability to the datasets. Moreover, these datasets provide recordings from multiple devices. In some cases, the devices were used to record the same acoustic scene. We thus removed such recordings to avoid duplicates (we only kept recordings from device A). 

The number of samples used from each dataset to build our final database is presented in \ref{tab:noiseDataset}: 
    
    \begin{table}[h!]
    \centering
    \resizebox{\columnwidth}{!}{
    \begin{tabular}{|c|c|c|c|c|}
    \hline
    Dataset                        & Number of acoustic scenes & Number of samples per class & Total & Notes                              \\ \hline
    DCASE 2017 development         & 15                        & 312                         & 4680  &                                    \\ \hline
    DCASE 2017 evaluation          & 15                        & 108                         & 1620  &                                    \\ \hline
    DCASE 2018 development mobile & 10                        & 864                         & 8640  & only device A is used              \\ \hline
    DCASE 2018 evaluation mobile   & 10                        & 1512                        & 15120 & only the first test folder is used \\ \hline
    DCASE 2020 development mobile  & 10                        & 1440                        & 14400 & only device A is used              \\ \hline
    DCASE 2020 evaluation mobile   & 10                        & 1188                        & 11880 & only the test folder is used       \\ \hline
    \end{tabular}
    }
    \caption{DCASE Datasets used for noise dataset construction} 
    \label{tab:noiseDataset}
    \end{table}

    \clearpage    
    This results in a final dataset containing \textbf{56340} audio samples from 20 different classes. However, this dataset is not balanced : certain classes are much more represented than others, which might be problematic for training a denoising model supposed to generalize well to any acoustic scene. For the training and validation datasets, we used a smaller but balanced dataset, consisting of 20 classes with 441 samples per class ( resulting in \textbf{8820} samples). We used samples from the \textit{2020 evaluation DCASE dataset} only to test the trained models. 

    The acoustic scenes belonging to both train and validation datasets are presented in the following table: 

    \begin{table}[h!]
    \centering
    \resizebox{\columnwidth}{!}{
    \begin{tabular}{|c|c|c|c|c|c|c|c|c|c|}
    \hline
    Airport & Beach  & Bus  & Café          & Car                                                         & \begin{tabular}[c]{@{}c@{}}City \\ center\end{tabular} & \begin{tabular}[c]{@{}c@{}}Forest \\ path\end{tabular} & \begin{tabular}[c]{@{}c@{}}Grocery \\ store\end{tabular} & Home  & Library \\ \hline
    Metro   & Office & Park & Public square & \begin{tabular}[c]{@{}c@{}}Residential \\ area\end{tabular} & Mall                                                   & Street pedestrian                                      & Street trafic                                            & Train & Tram    \\ \hline 
    \end{tabular}
     }
    \caption{Acoustic Scenes of our Background Noise Dataset}
    \end{table}
    
    \subsubsection{Selecting a Room impulse reponse dataset:}

    In order to simulate room reverberation, we used a dataset of real room impulse responses.  

    A room impulse response corresponds to the time domain response of a system ( microphone in a room ) to an impulsive stimulus. It mostly depends on the geometry and the material composition of the room, as it is due to the direct and indirect reflections in the room as well as the decay characteristics of the walls. Real IR are measured using \textit{Exponential Sine Sweep} or \textit{Maximum Length Sequence} techniques (\cite{DataAugmentationThesis}).

    We selected the \textbf{MIT IR Survey} \cite{MIT_IR_survey} dataset, which consists of \textbf{271 room Impulse Responses} measured in distinct locations. \footnote{A list of real room impulse responses datasets is given in this page :  \url{https://github.com/RoyJames/room-impulse-responses}}.

    \begin{figure}[H]
    \centering
    \begin{minipage}{0.45\textwidth}
    \centering
    \captionsetup{justification=centering}
    \includegraphics[width=0.9\textwidth]{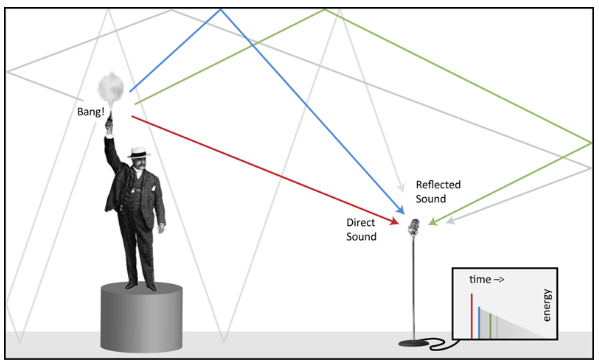}
    \caption{Direct and Indirect Reflections in a Room}
    \label{fig:reflections}
    \end{minipage}\hfill
    \begin{minipage}{0.45\textwidth}
        \centering
        \captionsetup{justification=centering}
        \includegraphics[width=0.9\textwidth]
        {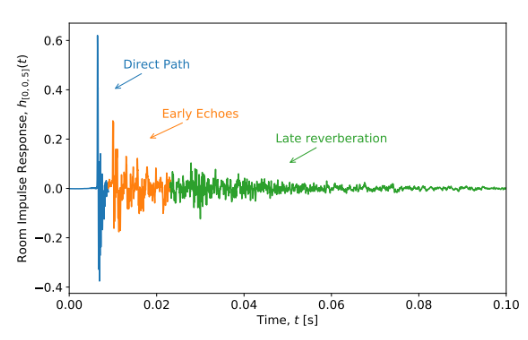}
        \caption{A measured Room IR}
        \label{fig:IR}
    \end{minipage}
    \end{figure}

    To select places, seven volunteers were asked at random moments of the day during 
    two weeks to respond with there locations. This allowed to obtain a diverse dataset, containing IRs of rooms that can be encountred in the course of daily life. We believe that this dataset presents enough variability to be used in our augmentation pipeline. 

    \subsubsection{Modeling Speakers and microphones:}\label{speakers_micro}

    We modeled speakers using the frequency responses of a collection of loudspeakers from the website : \url{https://pierreaubert.github.io/spinorama/}. 

    This website contains statistics about Loudspeakers. In particular, it contains their measured frequency responses, as well as their \textbf{-3db cutoff frequency} :

    \begin{figure}[H]
    \centering
    \begin{minipage}{0.45\textwidth}
    \centering
    \captionsetup{justification=centering}
    \includegraphics[width=1\textwidth]{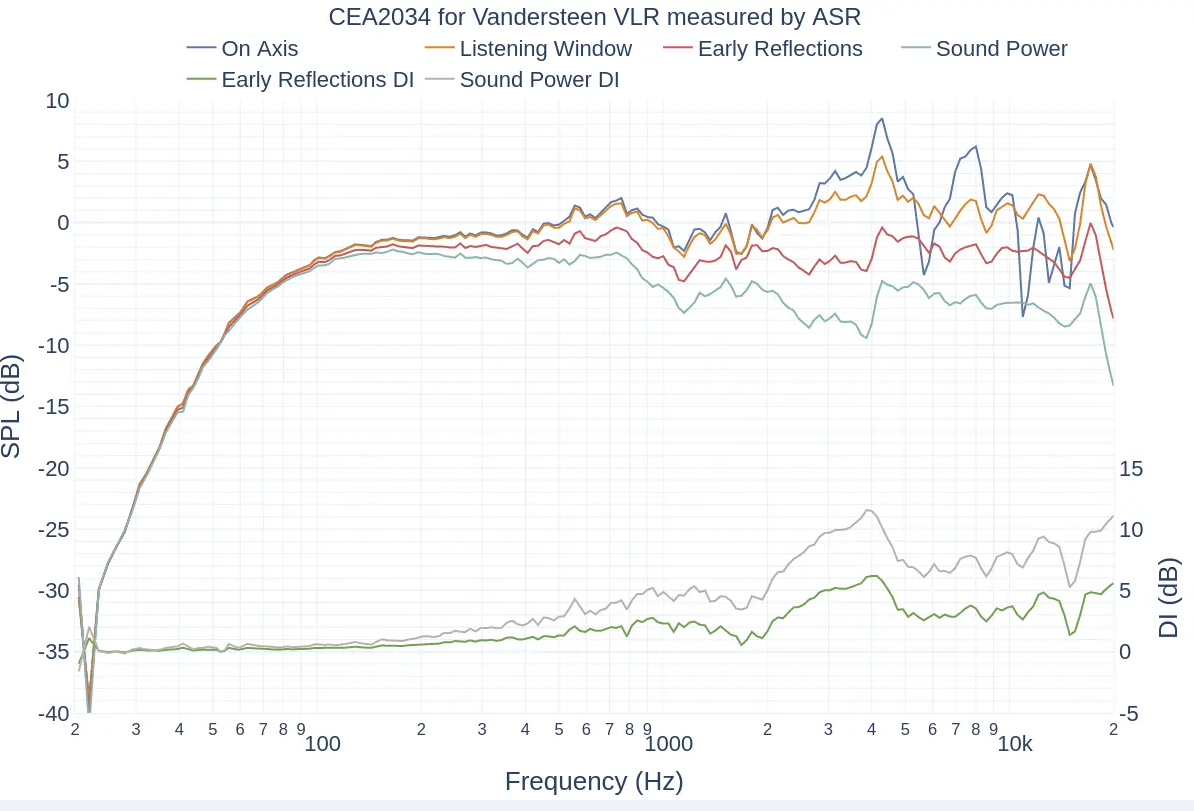}
    \caption{A Loudspeaker Typical Frequency Response (On axis curve)}
    \label{fig:freqResponse}
    \end{minipage}\hfill
    \begin{minipage}{0.45\textwidth}
        \centering
        \captionsetup{justification=centering}
        \includegraphics[width=1\textwidth]
        {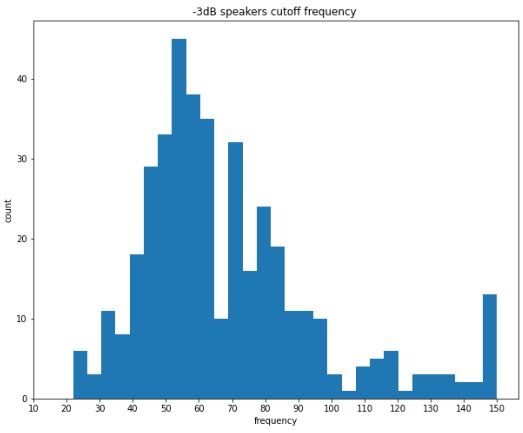}
        \caption{Spinorama Speakers -3db Cutoff Frequency Histogram}
        \label{fig:Histogram}
    \end{minipage}
    \end{figure}

    By looking at the appearance of the speakers frequency responses, as in \ref{fig:freqResponse}, we noticed that speakers could be modeled in first approximation with first order high and low pass filters. 
    
    We extracted the -3 db cutoff frequencies of \textbf{431 loudspeakers} using web-scrapping methods, which are presented in the histogram \ref{fig:Histogram}. This histogram shows that most speakers cutoff frequencies are located between 20 and 150 Hz. 

    Similarly, in the absence of a comprehensive review on smartphone microphone responses, we choose to approximate microphone distortion by applying low and high pass filters in typical frequency ranges, as given by some sources \footnote{\url{https://blog.faberacoustical.com/wpblog/2010/ios/iphone/iphone-4-audio-and-frequency-response-limitations/}}.

    \subsubsection{Constructing the audio augmentation pipeline : }

    The final data augmentation pipeline was built by adapting implemented transformations and pipelines from \textbf{torch-audiomentation} (\footnote{\url{https://github.com/asteroid-team/torch-audiomentations}}). This library contains a certain number of transformations that can be applied sequentially to raw audios waveforms, given in the form of pytorch tensors. In particular, it allows to specify the probability with which a transformation should be applied or not to each audio, as well as the parameters of each transformation. 

    The applied transformations are chosen based on \cite{DataAugmentationThesis}, \cite{MusicEnhancement} and \cite{NeuralAudioFingerprint}. The range of parameters of each transformation is chosen by listening to generated audios and by making them look as realistic as possible. The pipeline is applied to high quality audios tracks from the Deezer catalog, sampled at 44 kHz, to be able to correctly judge the influence of the chosen parameters on the augmented samples. 

    The final augmentation pipeline is the following \ref{fig:NoiseSoures}: 

    \begin{figure}[H]
    \centering
    \includegraphics[width=0.7\textwidth]{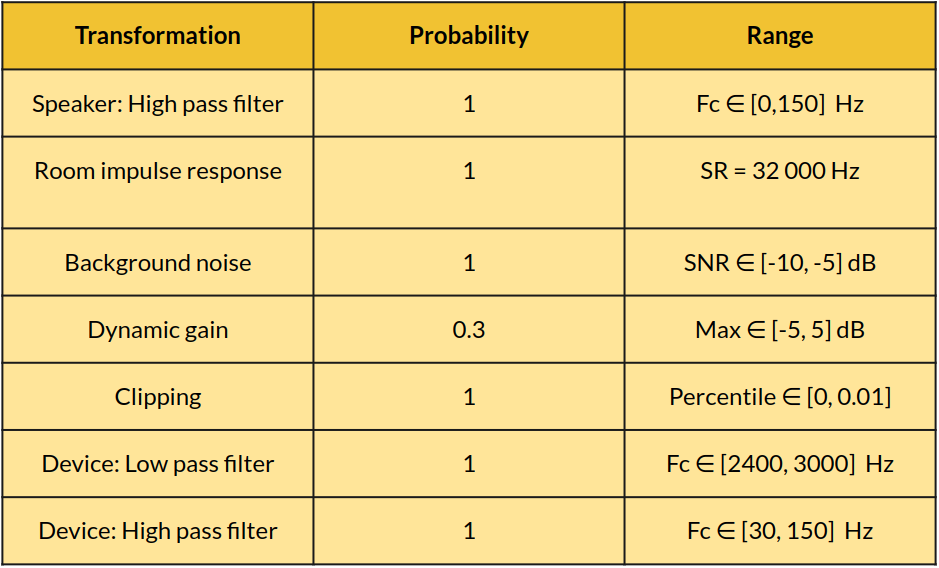}
    \noskipcaption{Final Augmentation Pipeline}
    \label{fig:NoiseSoures}
    \end{figure}

    For the background noise, we applied a signal to noise ratio between -10 and -5 dB, making the noise louder then the music signal. In addition to the speaker high pass filters, the room reverberation and background noise, we also added several transformations which represent potential degradation induced by recording devices: 

    \begin{itemize}
        \item \textbf{Gain} : during the recording, the signal can be distorted by variations in velocity. We decided to model this phenomenon using important gain distortions ( between -5 and 5 dB). 

        \item \textbf{Clipping} : artifacts can occur during recording and some samples can be lost. We artificially remove between 0 and 1 \%  of the audio samples randomly. 

        \item \textbf{Low pass and high pass filters} : The recording device has also a frequency response, which we model here by a combination of low pass and high pass filters, as explained in \ref{speakers_micro}. 

    \end{itemize}

   In practice, the noise constraints applied by this augmentation pipeline are strong. It requires an attentive listening to be able to identify the music in each recording. This is done to simulate real difficult cases faced by applications such as \textit{SongCatcher}.

    \section{Constructing a dataset with great musical variety}\label{dataset}

    We wanted to train a model on a dataset with great musical variety, to mimic a commercial catalog. Our model must be able to generalize well and denoise as many music pieces as possible. We selected a dataset developed by the Deezer research team that contains a large number of different musical genres : \textbf{the trackverse dataset}. 

    Each music genre contains its own musical characteristics in term of rythm, harmony, instruments and vocals. We believe that it is indeed important to train a model able to recognize the specificities of different music genres for better denoising. 
    
    This dataset consists of \textbf{48873} audio tracks from \textbf{25 different music genres} \ref{fig:MusicGenres}: 

    \begin{figure}[H]
    \centering
    \includegraphics[width=0.3\textwidth]{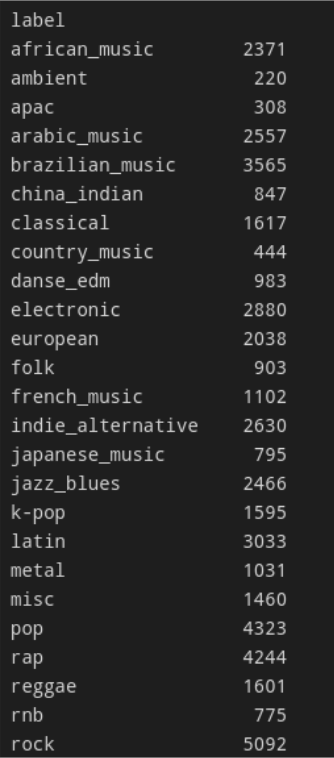}
    \noskipcaption{Music Genres of the Trackverse Dataset}
    \label{fig:MusicGenres}
    \end{figure}

    We extracted \textbf{10 second audio segments} from each track. The dataset was split in \textbf{train - val - test } using \textbf{80-10-10} proportions. We made sure that each split had its label proportion preserved. 

    Each audio fragment was processed by the noise augmentation pipeline at 44 kHz to generate clean-noisy audio pairs. We made sure that the audios from the \textit{test} dataset were augmented with a different background noise dataset then the one used in \textit{train} and \textit{validation}, to verify that the model can restore audios from different noise distributions.   

    We down-sampled both clean and noisy pairs to \textbf{5512 Hz} and generated their corresponding CQTs (so that they have the same configuration as in RADAR). We finally stored the 48873 clean-noisy CQTs pairs, constituting our final database. Exemples of noisy-clean CQTs pairs are given in \ref{fig:Clean-NoisyPairs2} and \ref{fig:clean-cqtPair}:

    \begin{figure}[H]
    \centering
    \includegraphics[width=1.05\textwidth]{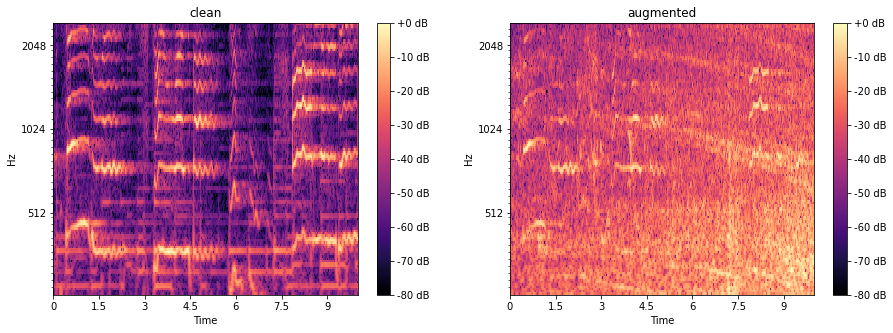}
    \noskipcaption{A Clean CQT and its Augmented Version  }
    \label{fig:clean-cqtPair}
    \end{figure}

    \section{Selecting potential Encoder-Decoder architectures}

    We selected 3 Encoder-Decoder architectures as potential candidates for the denoising model. The selected architectures are the following : 

    \begin{itemize}
        \item \textbf{Original U-net}, adapted from \cite{U-net}. The U-net architecture has been widely used in speech enhancement literature, as explained in \ref{SE}, as well as for denoising hitorical music recordings \ref{MusicDenoising} and performing music source separation \cite{Spleeter}. It has often reached state-of-the-art performances in these fields. It therefore constitutes a great candidate for our the denoising model.  

        \item \textbf{MMB-AIAT} : we wanted to adapt and test a more recent architecture from the speech enhancement literature. In 2022, the dual-branch attention in attention transformer \cite{AIAT} is one of the state of the art model in SE as we can see in \ref{fig:SOTA_AIAT}. The AIAT model is designed to process both magnitude and complex spectrogram, but can be adapted to only be trained on the magnitude component. The AIAT layer allows to capture long range dependencies along both time and frequency axes and at the same time aggregates global hierarchical contextual information. In this project, we relied only on the magnitude masking branch (MMB-AIAT) to process the noisy CQTs. The \textbf{encoder} consists of a densely-connected convolutional and an attention in attention transformer. 

        The densely convolutional encoder is composed of two 2D convolutional layers, followed by layer normalization (LN) and parametric ReLU (PReLU) activation. A dense-net with four dilated convolutional layers is employed between the two CNNs. The AIAT module consists of four adaptive time-frequency attention transformer based (ATFAT modules) and an adaptive hierarchical attention (AHA) module. The ATFAT modules can capture long range temporal-frequency dependencies, while the AHA module aggregate different intermediate features to capture multi-scale contextual informations. The \textbf{decoder} consists of the same dilated dense block as the encoder, followed by a sub-pixel 2D convolution module used to upsample the compressed features.  More details about the architecture can be found in \cite{AIAT}.

        \begin{figure}[H]
        \centering
        \includegraphics[width=1.0\textwidth]{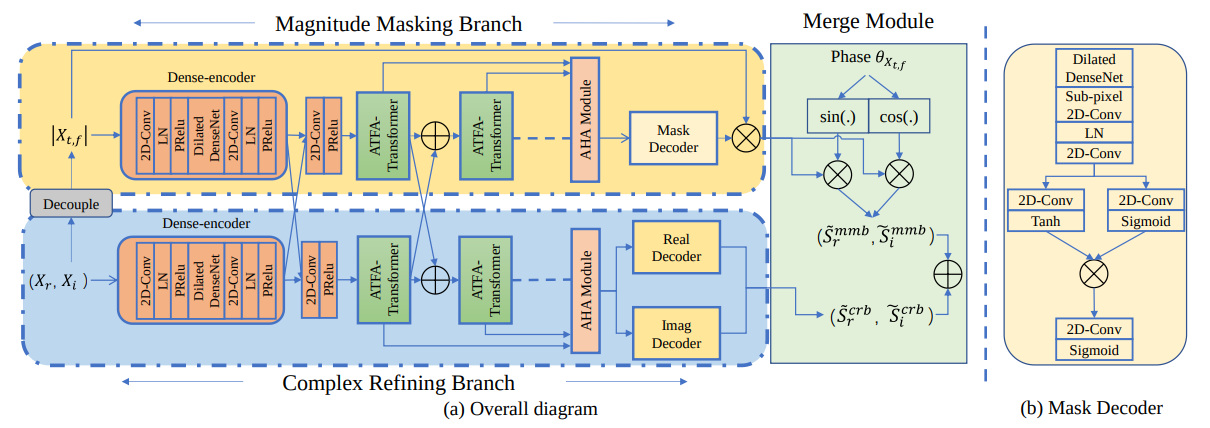}
        \noskipcaption{Dual Branch - Attention In Attention Transformer}
        \label{fig:AIAT}
        \end{figure}

        One advantage of this model is that it has few parameters (only 0.9M for the magnitude version), making it quite light and a potentially great candidate for our application.

        \item \textbf{Pix2pix} : this model corresponds to the generator used in the \textit{Music Enhancement via Image Translation and Vocoding} \cite{MusicEnhancement} paper. It is based on a ResNet generator adapted from Justin Johnson's neural style transfer project \cite{JohnsonNeuralStyleTransfer}. It has shown great performances for denoising music in \cite{MusicEnhancement}. This is the reason why we selected it : we wanted to test a model that already demonstrated abilities to process music and not only speech, as is the case for MMB-AIAT.  

        The architecture consists of 2 downsampling blocks (each containing 2D convolutional layers with kernel size 3 and stride 2, instance normalization and ReLu), 3 ResNet blocks (with kernel size 3, instance normalization) and two upsampling blocks (convtranspose2D with kernel sie 3 and stride 2, instance norm, ReLu). 
    \end{itemize}

    We focused on models that were fast or light. The following table \ref{tab:InferenceTime} present the inference time and number of parameters of the different models studied : 

    \begin{table}[h!]
    \resizebox{\columnwidth}{!}{
    \begin{tabular}{|l|c|c|c|}
    \hline
    \textbf{Model architecture} & \multicolumn{1}{l|}{\textbf{Inference time on cpu}} & \multicolumn{1}{l|}{\textbf{Inference time on GPU}} & \multicolumn{1}{l|}{\textbf{Number of parameters (in M)}} \\ \hline
    Unet                        & 748 ms                                              & 18.6 ms                                             & 31.03                                                     \\ \hline
    MMB-AIAT                    & 6073 ms                                             & 273.3 ms                                            & 0.91                                                      \\ \hline
    Pix2pix                     & 720 ms                                              & 19.79 ms                                            & 1.92                                                 \\ \hline
    \end{tabular} }
    \caption{Number of Parameters and Infernce Time of the selected Architectures}
    \label{tab:InferenceTime}
    \end{table}

   The inference time is estimated on 10 seconds CQTs (with RADAR configuration) of shape $(\textit{t}, \textit{f}) = (862, 117)$. To estimate the inference time on CPU, we fixed the number of CPU threads used by pytorch to two to simulate the fact that the model can run on low quality mobile devices or on limited server resources. For better precision results, we computed the average inference time over 100 samples on CPU and 10 samples on GPU.  \footnote{Note : We used \textit{torch.inference\_mode}, \textit{torch.cuda.Event} and \textit{torch.cuda.synchronize} to compute the inference times.}

    Both the U-net and Pix2pix model have a very low inference time : they require both only around 0.7s on a 2 thread CPU to process a 10 second CQT. This means that these models comply with our initial constraints. These models are fast because they are mainly constitued of convolutional layers. The MMB-AIAT model, in contrast, is quite slow and requires around 6s on the same CPU to process the CQTs, making it probably unsuitable for our application. This is mostly due to the use of attention layers, which can be quite slow (\cite{AttentionIsSlow}).  

    Concerning the number of parameters, the U-net model is much larger then the two others, with more then 30M parameters. This might be problematic for its integration in \textit{SongCatcher} depending on if the model runs on a server or a mobile device. Nevertheless, the 3 architectures have a rather low number of parameters in comparison to other models : the Demucs architecture  for SE \cite{Demucs}, for example, has more than 128M parameters.  
   
    \section{Defining relevant metrics for evaluation }

    It is important to define relevant metrics for both training monitoring and model evaluation to highlight the usefulness of the denoising model regarding the AFP system.  

    We are interested in two aspects : 
    
    \begin{itemize}
        \item We want first to show that the model is capable to \textbf{denoise the CQTs}, making them as close as possible to their clean references. 
        \item We also want to show that the model helps the AFP system with \textbf{spectral peaks preservation} : spectral peaks extracted by the AFP system from the denoised CQT should be as similar as possible to the ones extracted from the clean references.   
    \end{itemize}

    \subsection{Metrics to evaluate denoising capacities}

    Different metrics are used to monitor the evolution of the model's denoising capabilities (during training) as well as to compare models between themselves and with baselines (during evaluation).   

    The following used metrics come from the image denoising literature \cite{LossFunctionsImageDenoising} : 

    \begin{itemize}
        \item \textbf{L1-L2 losses} : we trained our models using L1 and L2 losses. As explained in \ref{losses}, these losses are often used in the denoising literature, and constitute also good metrics for measuring the distance between a noisy image and its clean reference. It corresponds to the sum of the absolute differences between predicted pixels and ground thruth pixels in the case of the L1 loss, and square error in the case of L2 loss.  
    \end{itemize}

    \begin{equation}
    \begin{split}
    \mathcal{L}_p(X_{noisy},X_{clean}) & = || {X_{noisy} - X_{clean}} ||_p \\
       & = \Big( \sum_{i,j} \big| X_{noisy}[i,j] - X_{clean}[i,j] \big|^p \Big)^{\frac{1}{p}}
    \end{split}
    \end{equation}

    \begin{itemize}
        \item The \textbf{PSNR} (peak signal to noise ration) is also commonly used to quantify images reconstruction quality \cite{PSNRvsSSIM}. The PSNR in dB, is given by : 
    \end{itemize}    

    \begin{equation}
    \begin{split}
        \text{PSNR} & = 10 \log_{10} \Big( \frac{MAX_I^2}{\mathcal{L}_1(X_{noisy},X_{clean})}\Big)\\ 
        & = 20 \log_{10} \big(MAX_I \big) - 10 \log_{10} \Big(\mathcal{L}_1\big(X_{noisy},X_{clean}\big)\Big)
    \end{split}
    \end{equation}

    where $MAX_I$ corresponds to the maximum possible pixel value of the images (255 for pixels coded on 8 bits).  

    \begin{itemize}
        \item The \textbf{SSIM} (structural similarity index) already introduced in \ref{losses}, can also serve as a metric. It compares images over windows of size $N*N$. The measure between two windows \textit{x} and \textit{y} is given by : 
    \end{itemize}

    \begin{equation}
      \text{SSIM}(x,y) = \frac{(2\mu_x\mu_y + C_1)(2\sigma_x\sigma_y + C_2)(2 \sigma _{xy} + C_3)} 
        {(\mu_x^2 + \mu_y^2+C_1) (\sigma_x^2 + \sigma_y^2+C_2)(\sigma_x\sigma_y + C_3)}
      \label{eq:SSMI}
    \end{equation}

    with $\mu_x$, $\mu_y$, $\sigma_x^2$, $\sigma_y^2$ are the average and variance of \textit{x},  \textit{y},  $\sigma_{x,y}$ is the covariance of \textit{x} and \textit{y}, and $C_1$, $C_2$, $C_3$ are two variables used to stabilize division with weak denominator. 

    More precisely, in this project, we used the Structural Dissimilarity (DSSIM) to measure the distance between noisy and cleans images :

    \begin{equation}
        \text{DSSIM} \big(X_{noisy}, X_{clean}\big) = \frac{1 - \text{SSIM}\big(X_{noisy}, X_{clean}\big)}{2}
    \end{equation}
    
    \subsection{Metrics to evaluate spectral peaks preservation}

     Even if the trained model is able to denoise the CQTs significantly, it doesn't necessarily mean ( or at least is doesn't explicitly demonstrate) that the model is increasing the AFP system's robustness to noise. Indeed, what we must show is that the denoising model helps the AFP system extracting  as much similar spectral peaks or landmarks as possible from the noisy CQTs and their corresponding cleans references. 
     
     We therefore defined some metrics to measure \textbf{spectral peaks preservation}. These metrics are based on the output of the last AFP peaks filtering step : it corresponds to the final extracted peaks obtained before landmark construction. We adapt the studied AFP system, RADAR, to get this output in the form of a \textbf{binary mask} satisfying the following constraints :    
     
     \begin{equation}
         M \big(i,j \big) = 
         \begin{cases}
            1 & \textrm{if (i\,j)} \in \textrm{S} \\
            0 & \textrm{otherwise} \\
         \end{cases}
     \end{equation}

     where $S$ corresponds to the set of spectral peaks coordinates. 

    \begin{figure}[H]
    \centering
    \includegraphics[width=1.0\textwidth]{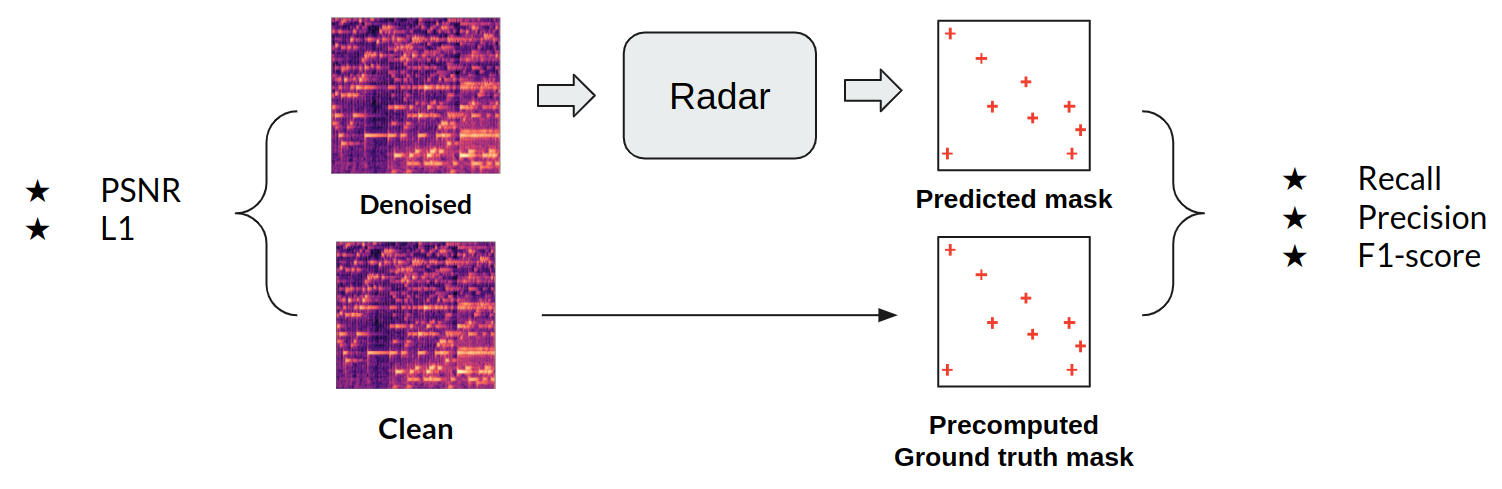}
    \noskipcaption{Metrics Computed on CQTs and their Corresponding Spectral Peaks}
    \label{fig:metrics_RADAR2}
    \end{figure}

    We compare the binary masks extracted by the AFP system from the denoised CQTs and their clean references using the Precision, Recall and F1-score (see below). These metrics rely on the notions of \textit{true positives}, \textit{false positives}, \textit{true negatives}, and \textit{false negatives}. Here,  a TP corresponds to a predicted spectral peak that is also present in the peaks ground truth mask associated to the clean music reference.  

    \begin{itemize}
        \item \textbf{Precision} : it corresponds to the percentage of retrieved peaks (extracted from the denoised CQT) that are also present in the original AFP (extracted from the clean reference).

        $$ \text{Precision} = \frac{\text{TP}}{TP + FP}$$

        \item \textbf{Recall} : it corresponds to the percentage of relevant peaks that have been retrieved, i.e the percentage of peaks from the clean CQT that are present in the denoised one  : 

        $$ \text{Recall} = \frac{\text{TP}}{TP + FN}$$

        \item \textbf{F1-score} : it corresponds to the harmonic mean between precision and recall. It is used as a metric to combine both precision and recall. 

        $$ F = 2 *  \frac{\text{precision}*\text{recall}}{\text{precision} + \text{recall}}$$
    \end{itemize}
    
    In RADAR,  the algorithm has a tolerance for spectral peaks position precision: it can consider a spectral peak from a query CQT identical to the one stored in the AFP database even if they have a difference of position of +/- one pixel in both time and frequency dimensions. 
    
    Therefore, for the previous three metrics, a predicted spectral peak is considered as a \textit{true positive}(TP) if it is within a window of size 3*3 around the spectral peaks of the clean reference CQT.      

    Even if we are computing both \textit{precision} and \textit{recall}, it is important to note that in this project, we are more interested in having a specific system with high precision, than a sensitive system with high recall. Indeed, we want to be confident on the extracted peaks, even if they are less numerous in the noisy CQTs than in its clean reference. Nevertheless, it is still important to keep a relatively high recall, otherwise there is a risk of extracting too few peaks. We will discuss in more detail the trade-off between precision and recall in the next sections. 
    
    \subsection{Computing metrics on intermediate AFP system steps}

    We generalized the computation of the different metrics to \textbf{intermediate AFP system steps} because these steps are more informative regarding spectral peaks preservation than the last RADAR step. The peaks in the last RADAR step are very sparse and can make metric scores quite low. Indeed, because of sparsity, each mispredicted peak has an important influence over the final metric scores and can make it decrease importantly. 

    On the opposite, intermediate AFP system steps contain much more information. These intermediate steps result in metrics with higher values and with are more discriminating: the gap in metric values between a good and a bad performing model is more noticeable. 
    
    In our study, we relied on the second and third RADAR steps to compute validation metrics: 

    \begin{itemize}
        \item  The second RADAR step, called \textbf{Octave energy}, corresponds to the last step before spectral peaks extraction. This step provides a sparse version of the CQT where a majority of the pixels have been discarded and set to 0, while most salient regions where kept. Images are perceptually informative : the main notes and the harmony of the music being played can still be observed. They can be compared using perceptual metrics from image denoising literature. In particular, we used \textbf{SSIM} and \textbf{L1}.
        
        \item The third RADAR step, called \textbf{Peaks extraction}, corresponds to the first step where spectral peaks are extracted. The peaks are then filtered in the following steps, reducing considerably their number. Computing the \textit{precision}, \textit{recall} and \textit{F1-score} of binary masks corresponding to the third RADAR step, both for clean and denoised CQTs, allows to obtain metrics that better highlight the performances of the models, while being still strongly correlated to preserving peaks of the overall AFP system. 
    \end{itemize}

    \begin{figure}[H]
    \centering
    \includegraphics[width=1.0\textwidth]{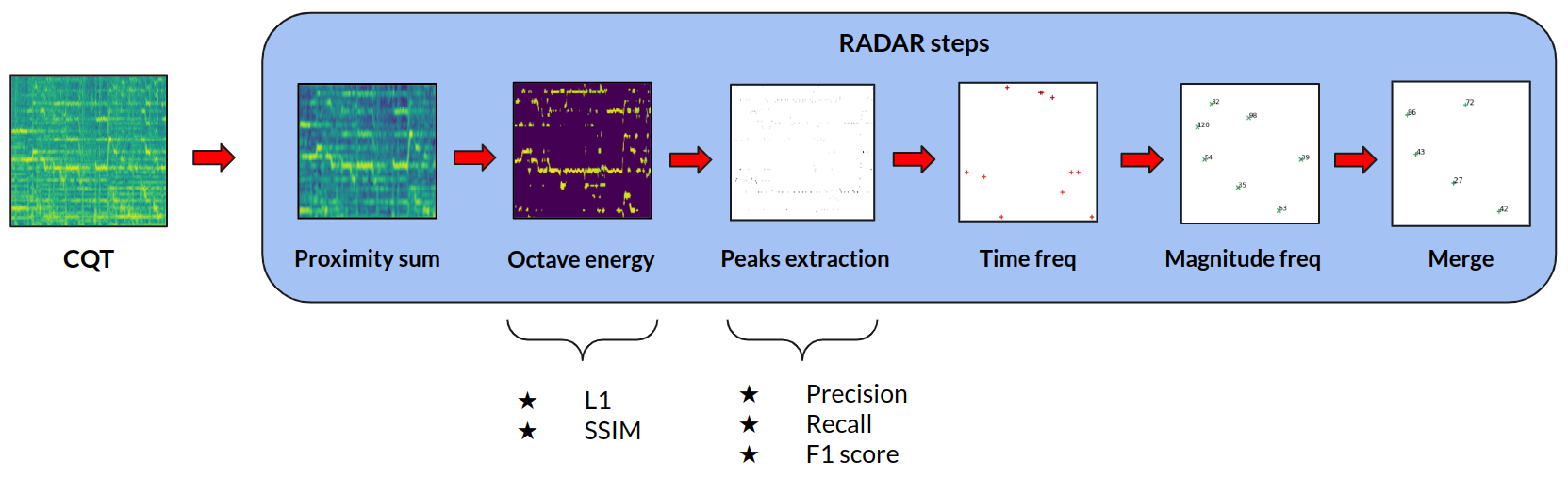}
    \noskipcaption{Metrics Computed on Intermediate RADAR Steps}
    \label{fig:metricsRadar}
    \end{figure}

    \chapter{Trainings}

    In this chapter, we present our main experiments, which aim at obtaining the best denoising model for our application. We trained models with various objectives:
    \begin{itemize}
        \item The first set of experiments aims at training a denoising model that it independent of the AFP system. We compare the performances of several trained architectures in terms of denoising capabilities and spectral peaks preservation, and select the one that better suits our application. 
        \item The second set of experiments aims at fine-tuning the selected architecture with a loss that improves the denoising model ability to preserve spectral peaks. This is done in order to increase the AFP's system specificity.   
    \end{itemize}
  
    \section{Learning to denoise}\label{Denoise}

    We trained the selected Encoder-Decoders to denoise the augmented CQTs. This was done in a supervised fashion : the models took as input the noisy CQTs and they were trained to minimize the mean absolute error (\textbf{L1 loss}) between the predicted CQTs and their clean references. The first trainings were done using the L2 loss, but we quickly noticed that the L1 loss allowed to obtain better performances regarding spectral peaks preservation. 

    We launched several trainings for each encoder-decoder architecture to fine tune the models. Each training was done using the same train and validation data set, which corresponds to the clean-noisy generated CQTs pairs from \ref{dataset}. The train set contains \textbf{39 098} clean-noisy CQTs pairs while the validation set \textbf{4887} pairs. 

    The CQTs used correspond to 10s audio extracts, but only 3s extracts were processed by the models for faster training. These 3s extracts are chosen randomly from each 10s CQT at each epoch. They thus vary from one epoch to another, increasing the number of examples seen by the models. We made sure that each model was trained with the exact same CQTs extracts by fixing the seed. 

    \subsection{Comparing Encoder-Decoder architectures}
    
    The following experiments were done using a \textit{batch size} of 8, \textit{adam optimizer} \cite{Adam} with variable learning rates and \textit{ReduceLROnPlateau} scheduler with patience 2 and 0.5 decreasing factor. We monitored losses and metrics on both train and validation data sets during training using weights and biases \footnote{https://wandb.ai/home}.

    The first experiments consisted in training the three different architectures using \textbf{L2 loss} and different \textbf{learning rates}. The goal was to find for which learning rate each model performed best as well as to compare their denoising and spectral peak preservation performances. The results of the different trainings are shown in \ref{fig:unet_lr_table} (UNet), \ref{fig:AIAT_table} (MMB-AIAT), and \ref{fig:pix2pix_table} (Pix2pPix).

    \begin{table}[H]
    \centering
    \includegraphics[width=1.0\textwidth]{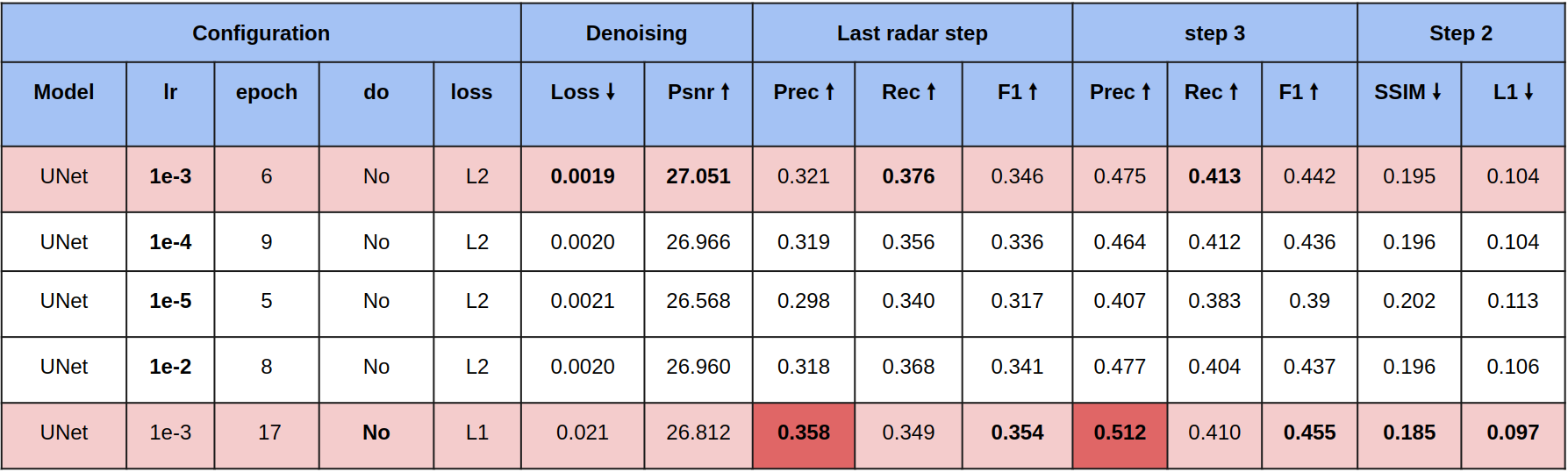}
    \noskipcaption{Learning Rate and Loss Influence over UNet, Results on Validation Set}
    \label{fig:unet_lr_table}
    \end{table}

    We trained the UNet using the following learning rates : \textit{1e-2, 1e-3, 1e-4, 1e-5}. As we can see in \ref{fig:unet_lr_table}, the best performances in the validation set were obtained for \textbf{LR = 1e-3} for all metrics : the PSNR reached 27.051 and precision and recall achieved 0.475 and 0.413 in RADAR step 3, 0.321 and 0.376 in the last RADAR step, showing better scores then the other UNets.

    Once the learning rate was selected, we added one experiment to see if the L1 loss leads to better results then the L2 loss. We observed that the L1 loss improved slightly the performance regarding spectral peaks preservation.
    
    Namely, with L1 loss, precision increased from \textbf{0.321} to \textbf{0.358} on the last RADAR step, and from \textbf{0.475} to \textbf{0.512} on the third RADAR step, while recall and PSNR remained rather stable. We also observed an important decrease in SSIM and L1 loss over RADAR step 2 images (passing from 0.195 to 0.185 and from 0.104 to 0.097 ). A similar behaviour was noticed on the \textit{pix2pix} model when trained on a L1 loss, as shown in \ref{fig:pix2pix_table}. 

    This is probably due to the fact that the L2 loss is more sensitive to abnormal points than the L1 loss, and to the fact that RADAR extracts peaks with high magnitude.
    The L2 loss has more difficulties to reconstruct points with high magnitudes because they might strongly increase the reconstruction error. This result in less extracted peaks and thus a lower precision.

    \begin{table}[H]
    \centering
    \includegraphics[width=1.0\textwidth]{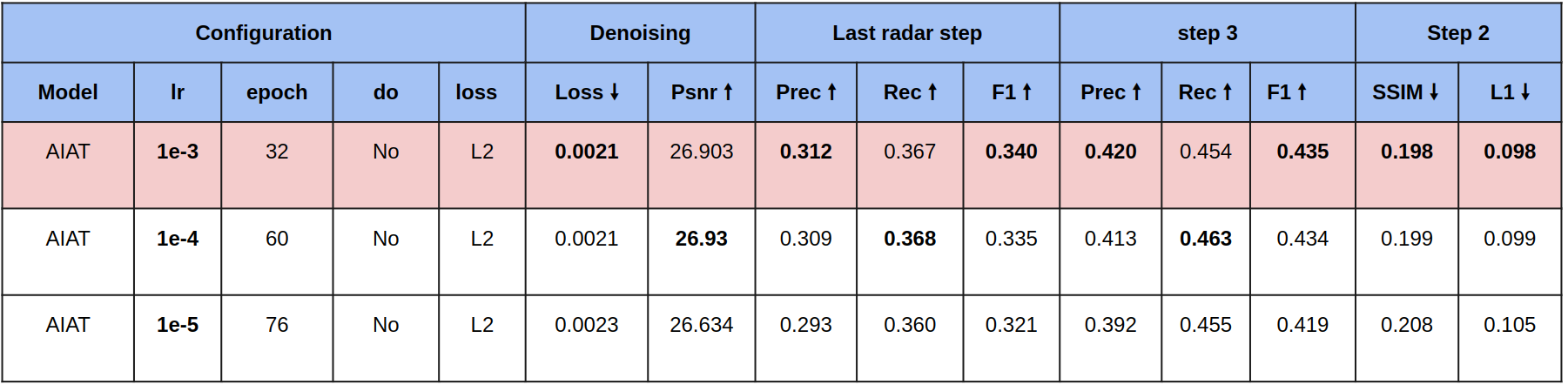}
    \noskipcaption{Learning Rate Influence over AIAT, Results on Validation Set}
    \label{fig:AIAT_table}
    \end{table}

    \begin{table}[H]
    \centering
    \includegraphics[width=1.0\textwidth]{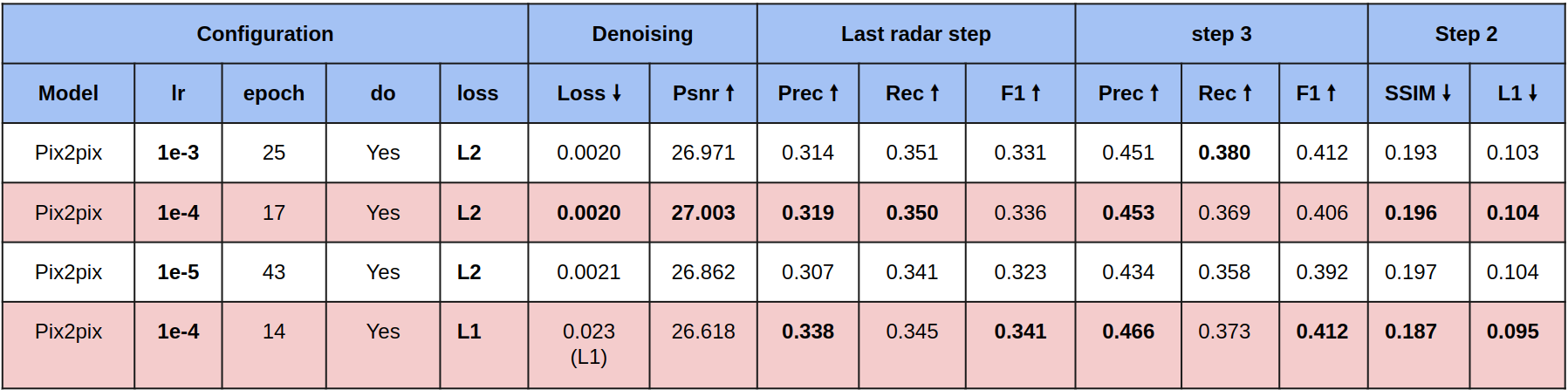}
    \noskipcaption{Learning Rate and Loss Influence over Pix2pix, Results on Validation set}
    \label{fig:pix2pix_table}
    \end{table}

    We also trained  both \textit{MMB-AIAT} and \textit{Pix2pix} models on different learning rates : \textit{1e-3, 1e-4, 1e-5}. We found that the best performance in precision was obtained with \textbf{LR = 1e-3} for \textit{MMB-AIAT} and \textbf{LR = 1e-4} for \textit{Pix2pix}, as we can see in \ref{fig:AIAT_table} and \ref{fig:pix2pix_table}. 

    However, in the end, these two models do not seem to perform significantly better than the UNet architecture. The best UNet model obtained the best performance in almost all metrics in comparison to the best pix2pix and MMB-AIAT models trained with the same loss (L2) \ref{fig:3models_table}: 
    
    \begin{table}[H]
    \centering
    \includegraphics[width=1.0\textwidth]{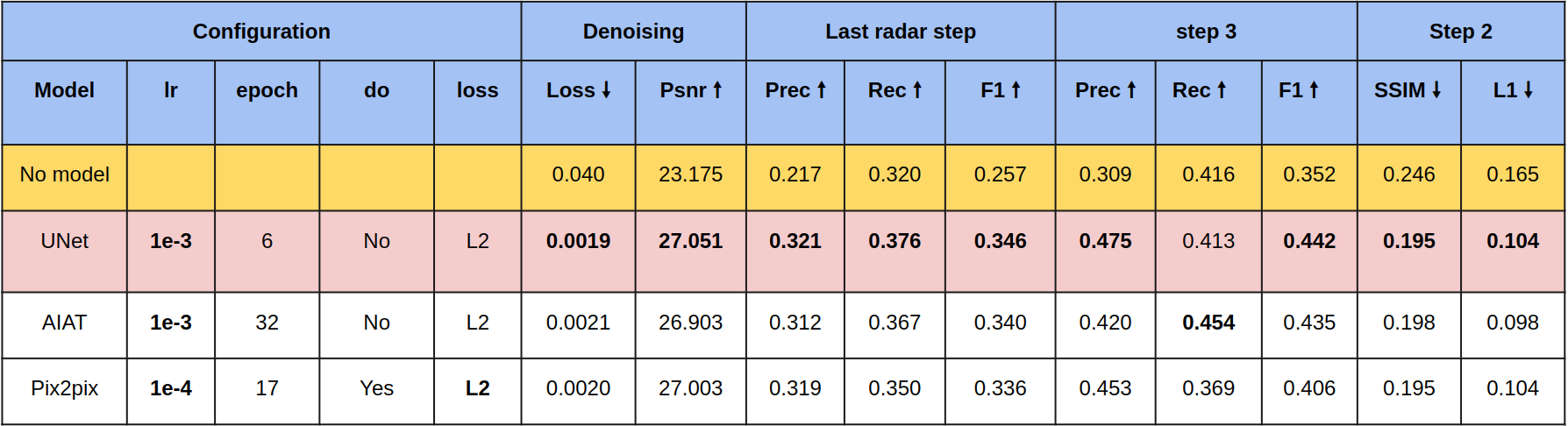}
    \noskipcaption{Comparison of Architectures and Baseline Performances}
    \label{fig:3models_table}
    \end{table}

    Finally, on the previous table \ref{fig:3models_table}, we compare the best models obtained for each architecture obtained with L2 loss to the values of the metrics in the case where no denoising model is integrated in the AFP pipeline. In the "No model" row, metrics are computed between the noisy and clean references CQTs, as if the noisy recording were directly processed by the RADAR pipeline, without passing through a denoising model. 
    
    We observe a \textbf{substantial improvement over all metrics between the "no model" baseline and the UNet model}. In particular, there is an important increase in PSNR thanks to the denoising model, as we go from a noisy CQT with \textbf{23.175} PSNR, to a denoised one with \textbf{27.051}. 

    Although the primary goal of the training was to learn to denoise and not to learn to preserve peaks, we observe that denoising the CQTs allows for a significant improvement over RADAR metrics. The most interesting result is the important increase in precision of the system, passing from \textbf{0.309} to \textbf{0.475} in the last RADAR step which \textbf{multiplies the system precision by more than 1.5}. We noticed however that the denoising model hardly improves the system recall, which is stagnant. 
    
    This is probably due to the behaviour of the denoising model : the models learns primarily to remove noise from the noisy CQT, eliminating spectral peaks that do not belong to the AFP of the clean CQT, which increases the precision. However, the algorithm struggles to generate "new" spectral peaks that are present in the clean CQT but not in the noisy one : it is hard to artificially reconstruct peaks that have been damaged by the additive noise. This is likely to be the reason why recall is stagnant.   

    In this experiment, we finally  select the UNet with learning rate 1e-3 and trained with loss L1 as the best candidate.

    \subsection{Finetuning the UNet with dropout }\label{dropout}

    We fine-tuned the selected UNet from the previous section using Dropout. Dropout can improve models performances, allowing them to better generalize and avoid overfitting \cite{DropOut}. Here, we study the influence of different dropout rates over the UNet architecture. 

    More precisely, we add dropout after each downsampling block of the encoder architecture, and we modify the dropout probability rate from an experiment to another. The results are presented in the following table \ref{fig:UNet_do_table} : 
    
    \begin{table}[H]
    \centering
    \includegraphics[width=1.0\textwidth]{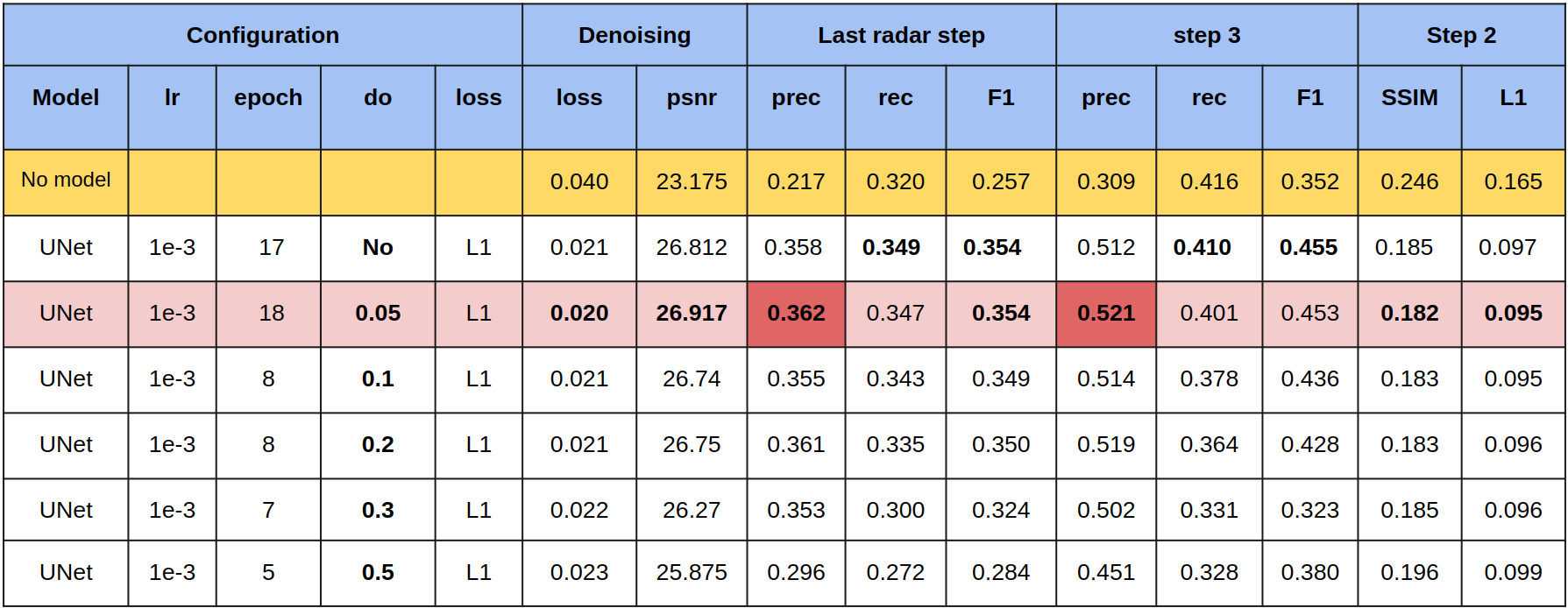}
    \noskipcaption{Dropout Influence over UNet, Results on Validation Set}
    \label{fig:UNet_do_table}
    \end{table}

    The best results are obtained with a \textbf{dropout rate p = 0.05}. We note, however, that the results of the UNet with no dropout are quite close. 

    The best model trained using the denoising framework reaches \textbf{a precision of 0.52 on the third RADAR step (vs 0.309 with no model)}, while the recall remains almost constant (0.401 for the UNet vs 0.410 with no model). 

\begin{figure}[H]
    \centering
    \includegraphics[width=1.0\textwidth]{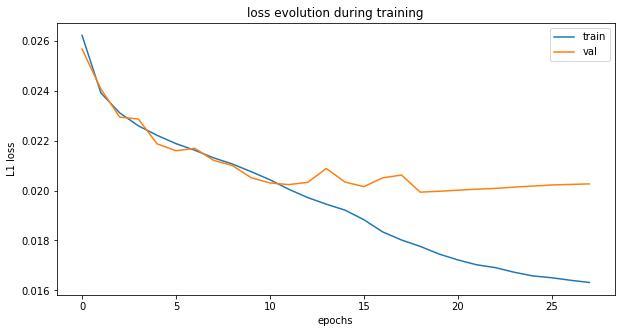}
    \noskipcaption{Loss Evolution during Training of the Best Denoising Model, on Train and Val sets}
    \label{fig:training_UNet}
    \end{figure}

    We display the loss evolution during training of both train and val datasets for the final selected UNet model in \ref{fig:training_UNet}. We observe that the validation loss no longer decreases after a certain number of epochs (around 11). The minimum in validation is reached at step 18, which corresponds to the final selected model. Training continues until epoch 27 but we only kept the model with best score on the validation set (we used early stopping with patience 10 ). 
    
\begin{figure}[H]
    \centering
    \begin{minipage}{0.45\textwidth}
    \centering
    \captionsetup{justification=centering}
    \includegraphics[width=1\textwidth]{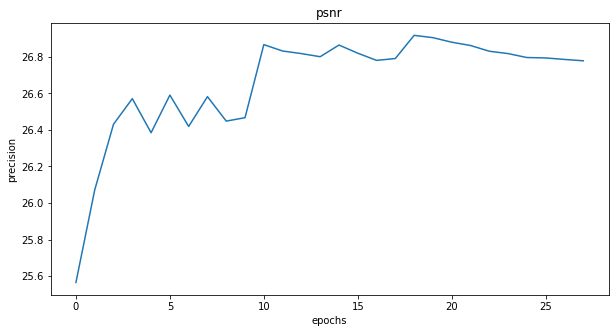}
    \caption{Evolution of the PSNR during Training, results on Val set}
    \label{fig:PSNR_evolution}
    \end{minipage}\hfill
    \begin{minipage}{0.45\textwidth}
        \centering
        \captionsetup{justification=centering}
        \includegraphics[width=1\textwidth]
        {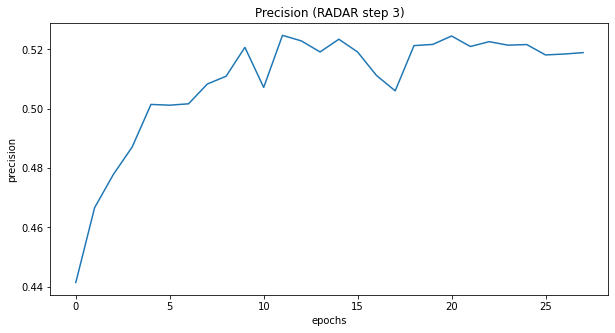}
        \caption{Evolution of the Precision during Training, results on Val set}
        \label{fig:PrecisionEvolution}
    \end{minipage}
    \end{figure}

    We also plot the evolution of the PSNR and precision on the validation set during training in \ref{fig:PSNR_evolution} and \ref{fig:PrecisionEvolution}. Here, we observe that the precision of the AFP system increases with its ability to denoise the CQTs, even though the training didn't focus directly on increasing the system's precision. This is not surprising : the more the CQTs will be restored (denoised), the more they will have common salient peaks with the clean CQTs that will be extracted. 
    
    \section{Visualising the denoised CQTs}

\begin{figure}[H]
    \centering
    \includegraphics[width=0.8\textwidth]{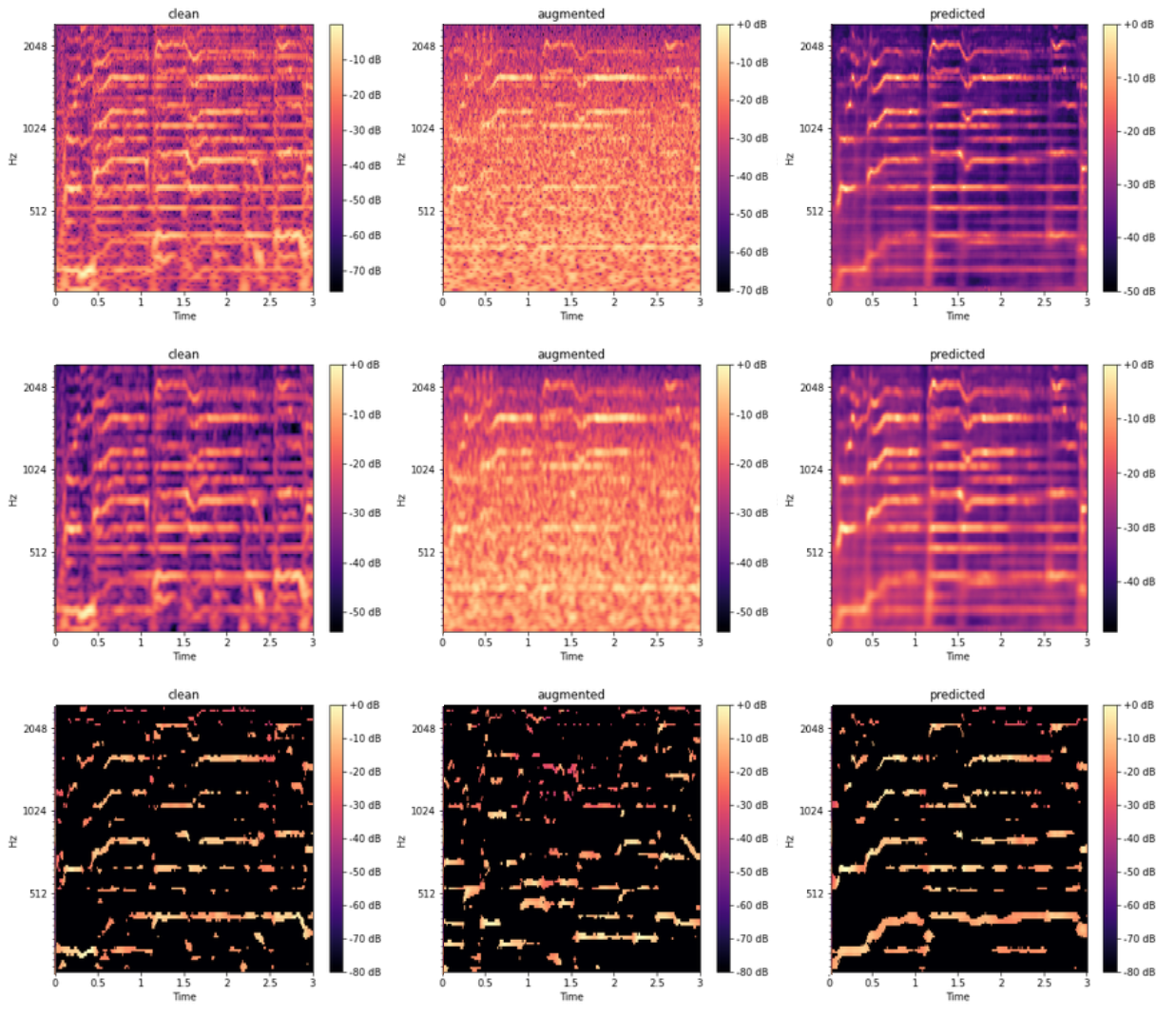}
    \noskipcaption{Clean, Augmented and Denoised CQTs }
    \label{fig:CAD_1}
    \end{figure}

    We visualize some denoised CQTs obtained with the denoising model selected in the previous part in \ref{fig:CAD_1}, as well as in the annex with \ref{fig:CAD_2} and \ref{fig:CAD_3}. In \ref{fig:CAD_1}, the denoised CQTs are compared to their clean references and their corresponding noisy versions. We also display the associated features of the first RADAR step (in the middle) and of the second RADAR step (in the bottom).

    In the previous figure, we can reasonably hypothesize that the denoising model has learned to remove noise from the noisy CQT. This has an impact on the different steps of the AFP system features. We can notice that the octave energy feature (in the bottom) associated to the \textit{predicted CQT} is quite similar to the one of the clean reference : the long lines (likely to be notes) present in the clean CQT have been also extracted from the predicted CQT, while it is not the case for the augmented one. This indicates that through denoising, the model has also helped the AFP system to extract more spectral peaks present in the clean CQT reference, increasing its precision. 
        
    \section{Learning to preserve peaks} \label{denoising}

    \subsection{Designing the Deep Radar Feature Loss}
    
    In the previous section, the model is trained to denoise the CQTs but does not directly learn to preserve spectral peaks. We want to train a model that learns how to extract from the noisy CQT as many spectral peaks as possible that also belong to the corresponding clean CQT.
    
    Our idea is to introduce a new loss that makes the model learn how to preserve spectral peaks. We took inspirations from the Deep Feature Loss \cite{DeepFeatureLoss} to develop what we call the \textbf{Deep Radar Feature Loss} \ref{fig:DRFL} : 

    \begin{figure}[H]
    \centering
    \includegraphics[width=1.0\textwidth]{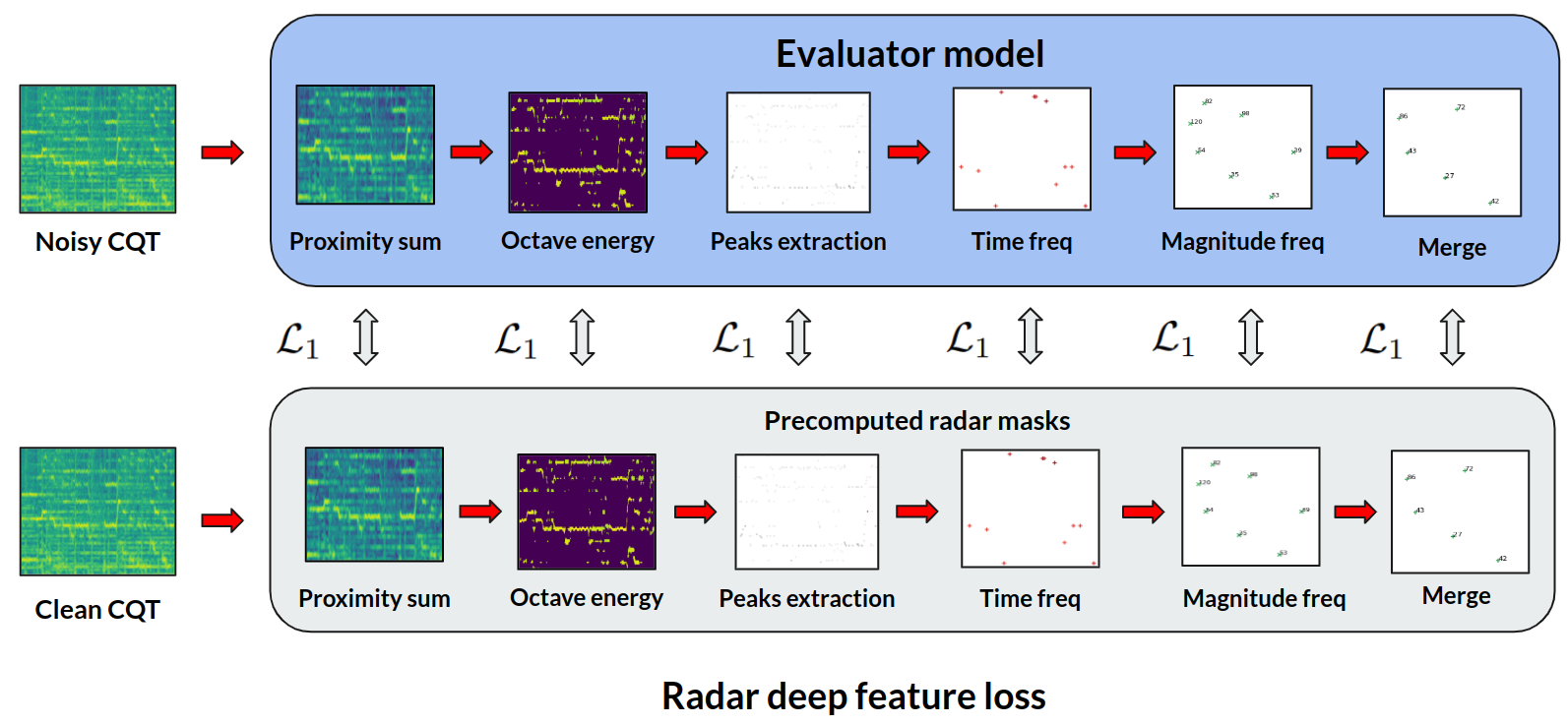}
    \noskipcaption{Deep Radar Feature Loss}
    \label{fig:DRFL}
    \end{figure}

    The principle of the Deep Feature Loss is explained in \ref{losses}: the idea is to compare the features of a pretrained DL model between a prediction and its ground thruth. We adapt this to RADAR: it can also be seen a model with fixed weights which serves as a feature extractor. Here, each step of RADAR can be seen as a feature of the CQTs. 
    
    The idea of the loss is then to compare the features of the clean and noisy CQTs and reduce their distance using an image comparison loss such as the L1 loss in \ref{fig:DRFL}. Since these features are directly used to extract spectral peaks in a deterministic way, we thought that reducing the distance of the noisy features to their clean reference would implicitly make the model learn how to preserve more spectral peaks from the clean CQTs. 

    In practice, the implementation of the \gls{DRFL} was a difficult task. Indeed, implementing a peak selection engine in a framework that enables gradient flow appears as quite challenging. In particular, RADAR has been developed without the differentiability constraint and with frameworks that do not account for gradient flow. This is problematic if we want to propagate the gradient of the distance between features to the model being trained. 
    
    We had thus to re-implement RADAR using Pytorch to allow for gradient flow. Because of the significant engineering efforts required by this task, we finally implemented only the three first RADAR steps (including Peaks extraction) with Pytorch. 

    We tested our loss for training and were confronted with several challenges. First, the training became very slow, because the different steps of the DRFL were quite heavy: the train epoch took about 30 minutes with only the L1 loss against 90 minutes with the DRFL (using the first 2 steps of RADAR). In particular, the third RADAR step made the training much too slow, making it unusable during training. Engineering optimisation could probably help reduce this time, but it is beyond the scope of this work. 

    We tested the DRFL with several configurations. We wanted to verify which features were useful to improve the model's performances regarding spectral peaks preservation. We made several observations : 

    \begin{itemize}
        \item Computing the L1 distance between several features (from steps 1, 2 and 3) at the same time made loss convergence difficult : the model is asked to optimize regarding multiple directions at the same time which makes it predict nonsensical outputs.
        
        \item Relying only on one feature during training resulted in better results. It was also easier to interpret the training curves evolution and understand the model training behaviours.

        \item Training the model using the second or third RADAR step showed interesting learning behaviours regarding spectral peaks preservation when tested on small datasets. In comparison, training the model using only the first step of the DRFL (\textit{Proximity sum}) didn't show any improvement regarding spectral peaks preservation.  

        \item Because training was too slow with the third RADAR step, we finally decided to focus on trainings with the \textbf{second step of the DRFL}. We thought it would be interesting to investigate how we could improve the model performances as much as possible by focusing on the fine-tuning and adaptation of this second step, while the engineering optimization of the DRFL is left for future works.    
    \end{itemize}

    During trainings with the DRFL, we first noticed that the minimization of the distance between the \textit{octave energy} (RADAR step 2) images of the clean and noisy CQTs during training could lead to strange results, with no more spectral peaks extracted from the noisy CQTs at the third RADAR step. This is due to the important sparsity of the octave energy masks and the fact that the L1 loss is not adapted for sparse images : during training, the model learns that in order to minimize the distance between clean and noisy octave energy masks, the best thing to do is to generate a mask with null values. 

    This led us to consider other losses, better suited for sparse image comparison. In particular, we were interested by the Tversky loss \cite{TverskyLoss} and the Focal Tversky loss \cite{FocalTverskyLoss}. These losses are well adapted for dealing with images with high class imbalances, and are often used for image segmentation tasks. 

    The Focal Tversky loss is based on the \textbf{Tversky index}, which is given by : 

    $$ TI = \frac{TP}{TP + \alpha FP + \beta FN }$$

    where TP, FN and FP correspond to the sum of \textit{true positive}, \textit{false negatives} and \textit{false positives} values, and $\alpha$ and $\beta$ are two parameters satisfying $\alpha + \beta = 1$

    In this project, we are not dealing with segmentation images as we are predicting continuous values and not discrete classes for each pixel of the image. For this reason, we had to slightly adapt the notions of TP, FP and FN. We considered here the problem as binary : a positive corresponds to a non null pixel (for which we keep its continuous value), while a negative correspond to a null pixel. In octave energy images, negatives are the majority class.  We redefined the notions of TP, FP and FN in the following way : 

    \begin{itemize}
        \item A pixel is considered as a \textit{true positive} if both the ground truth and predicted octave energy images have non zeros values at the corresponding pixel coordinates. The associated true positive value is then given by : $pixel_{GT} * pixel_{pred}$
        \item A pixel is considered as a \textit{false positive} if the value at the corresponding coordinates is non null in the predicted octave energy image but null in the ground truth. The false positive value is then given by : $pixel_{pred}$
        \item A pixel is considered as a \textit{false negative} if the prediction's value is null but the ground truth is non null at the corresponding coordinates. The false negative value is then given by : $pixel_{GT}$
        \item Finally, a \textit{true negative} corresponds to both GT and prediction having a null value at the same coordinates. We do not associate a value to true negatives as they are not involved in the Tversky index.  
    \end{itemize}

    The Tversky loss is given by $(1 - TI)$. The goal of this loss is to minimize the FN and FP values by making the TI index as close as possible to one. The parameters $\alpha$ and $\beta$ allow to control how much we want to penalise FN over FP : by setting $\alpha > \beta$, we want to penalise false positive more. In practice, it is common to test several parameters and find which one are best suited for the considered problem. 

    The Focal Tversky loss is a generalization of the Tversky loss. It is given by : 

    $$ FTL = \big( 1 - TI \big )^{\gamma}$$

    where $\gamma$ is a parameter that controls the non linearity of the loss. 

    The non-linear nature of the loss gives us control over how the loss behave for different values of the \gls{TI}. When $\gamma > 1$, the model is forced to focus on hard examples, as the loss gradients are higher for examples where $TI < 0.5$. When $\gamma < 1$, the model focuses on examples where $TI > 0.5$. This is useful to continue to learn near convergence. In practice, the best parameters $\alpha$, $\beta$ and $\gamma$ are usually determined manually. In \cite{FocalTverskyLoss}, the values $\alpha = 0.7$, $ \beta = 0.3$ and $\gamma = \frac{3}{4}$ have shown the best results in precision and recall  ( see \ref{fig:FTL_table} \footnote{Note that the paper uses $\frac{1}{\gamma}$ instead of $\gamma$}). 

    In this project, we used the Focal Tversky Loss applied to the second step of the DRFL to train the UNet architecture. We observed that the model had trouble converging when trained from scratch with the DRFL. However, fine-tuning the UNet obtained in \ref{dropout} for a few epochs with the DRFL, using a small learning rate, allowed us to obtain important improvements in precision. 

    \subsection{Fine-tuning with the DRFL}

    We trained the model with several learning rates and observed that the FTL converged only when very small learning rates were used (in the order of \textit{1e-7, 1e-6}).  
    
    We also tested the influence of different  $\alpha$ and $\beta$ parameters on training, that are presented in \ref{fig:DRFL_table}. For these experiments, we used a batch size of 4. It is smaller than the one used for the denoising experiments in \ref{denoising} because of the DRFL which requires higher GPU resources. We also fixed the maximum number of epochs to 20 and reduced the early stopping patience to 5. Indeed, we observed that after a certain number of epochs (around 10), performances didn't improve much. We also set the same number of epochs for each run in order to better compare the influence of the $\alpha$ and $\beta$ parameters over model performances.     

    \begin{table}[H]
    \centering
    \includegraphics[width=1\textwidth]{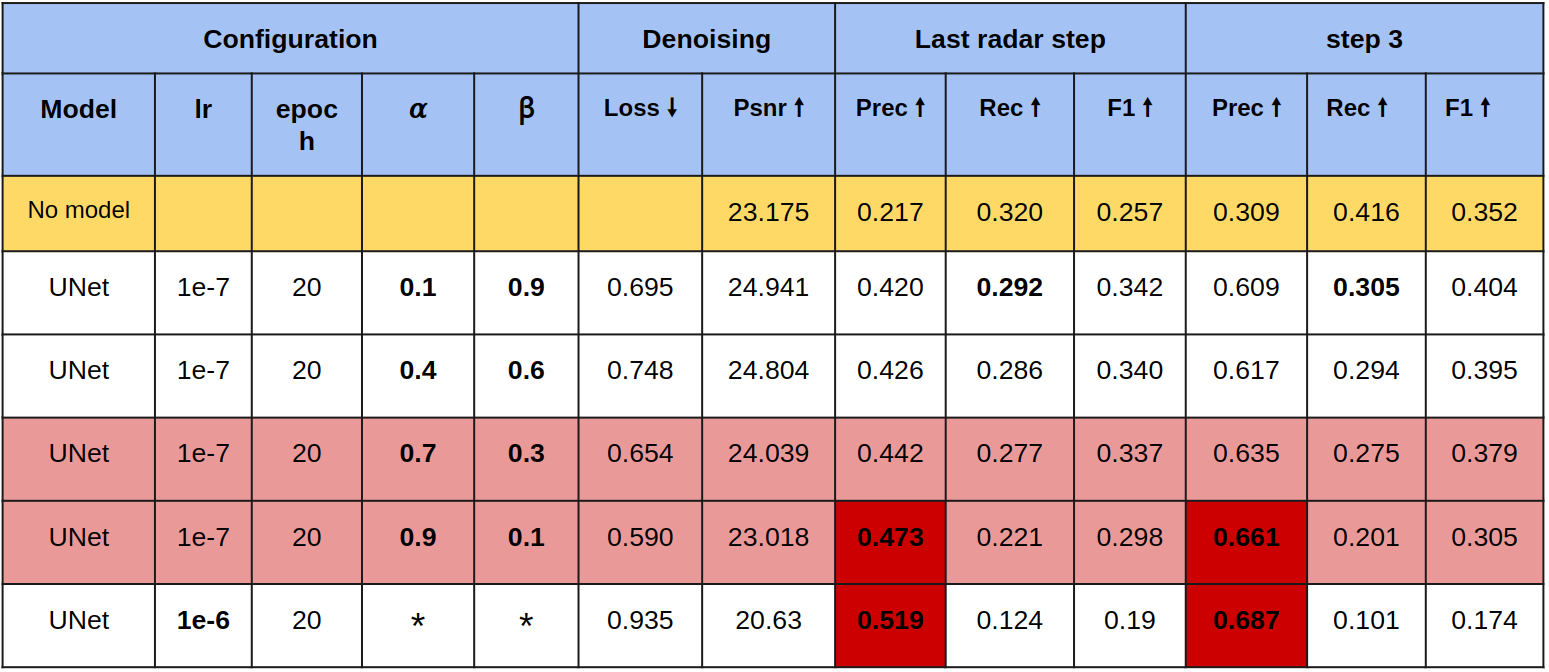}
    \noskipcaption{Influence of $\alpha$ and $\beta$ Parameters over UNet Performances, Results on Validation Set}
    \label{fig:DRFL_table}
    \end{table}

    Fine-tuning the UNet trained with L1 loss with the DRFL allows to considerably improve the precision performances of the model. In \ref{dropout}, the best UNet model obtained a precision score of 0.36 on the last RADAR step and 0.521 on the third RADAR step. Here, the model with highest precision reached \textbf{0.51} on the last RADAR step and \textbf{0.68} on the third RADAR step. This score is more than two times higher than the precision of the AFP system when no model is present (0.309 vs 0.687 for the third RADAR step).  

    However, this increase in precision is accompanied with an important decrease in recall, reaching 0.124 on the last RADAR step (vs 0.32 with no DL model).  Even if we are interested in having a model with high precision and that we can be more tolerant in terms of recall, having an extremely low recall indicates that the model behaves badly : it no longer extract a significant number of spectral peaks. The system extracts spectral peaks for which it is confident but as a consequence is not extracting enough of them. This results in a F1-score that is even lower than the F1 score of the system with no DL model (\textbf{0.19} vs \textbf{0.25} for the last RADAR step). Poor results in term of recall were obtained when training models with a learning rate of $1e-6$. In this setting, the model converges too quickly, increasing importantly its precision. With this learning rate, the same behaviour was observed for different configurations of the parameters $\alpha$ and $\beta$.

\begin{figure}[H]
    \centering
    \begin{minipage}{0.5\textwidth}
    \centering
    \captionsetup{justification=centering}
    \includegraphics[width=1\textwidth]{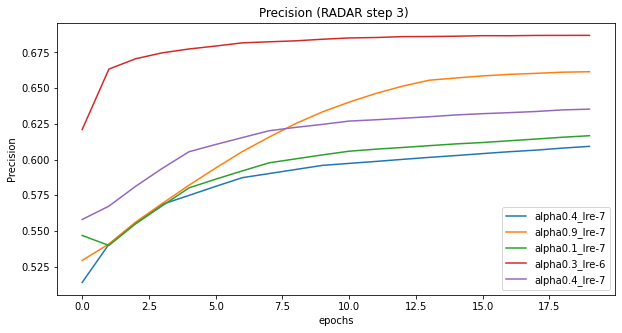}
    \caption{Evolution of the Precision during Training, Results on Validation Set}
    \label{fig:precisionRadar3}
    \end{minipage}\hfill
    \begin{minipage}{0.5\textwidth}
        \centering
        \captionsetup{justification=centering}
        \includegraphics[width=1\textwidth]
        {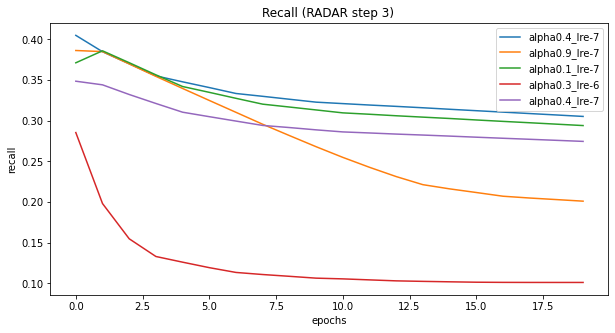}
        \caption{Evolution of the Recall during Training, Results on Validation Set}
        \label{fig:recall_radar3}
    \end{minipage}
    \end{figure}

    We observed more interesting results when we decreased the learning rate to $1e-7$. With such a small learning rate, we could clearly see the influence of the hyper-parameters $\alpha$ and $\beta$ on performances : as we can see in \ref{fig:DRFL_table}, they ensure a trade-off between precision and recall. When we increase $\alpha$ (and thus decrease $\beta$), we increase the precision of the system but decrease its recall. This is particularly visible on the figures \ref{fig:precisionRadar3} and \ref{fig:recall_radar3}, where the opposite behaviours of curves in the two figures is observed. The more a model reachs a high precision in \ref{fig:precisionRadar3}, the more it will have a low recall in \ref{fig:recall_radar3}. 

    This behaviour is due to the role of $\alpha$ and $\beta$ in the FTL. When we increase $\alpha$, we penalise more \textit{false positives} : we force the model to be more confident on the spectral peaks extracted, which increases its specificity (or precision). When we decrease $\beta$, we are also penalizing less \textit{false negatives}, encouraging the model to not extract spectral peaks (giving null values in the octave energy image) when it is not confident about them, decreasing thus the recall.     

    Selecting the right parameters $\alpha$ and $\beta$ allow us to increase importantly the specificity of the system while trying to not affect to much its sensitivity. In particular, here, we consider that models with a high precision and an overall F1-score that is higher than the F1 score of "no-model" constitute potential great candidates to be integrated in the final AFP system. However, it would require more investigations to see exactly how the precision and recall of the system would impact its identification performances. 

    The final model trained with the DRFL that we select as the "best models" for spectral peaks preservation are the models with parameters $\alpha = 0.7$ and $\alpha = 0.9$. These models reach a precision of 0.442 and 0.473 on the last RADAR step, 0.635 and 0.687 on the third RADAR step, \textbf{doubling the precision of the system over the "no-model" baseline (precision of 0.217 and 0.309 in each RADAR step), while their final F1-score is still over the baseline} ( 0.337 and 0.298 on the last RADAR step vs 0.257).

    We note however that fine-tuning the UNet with the DRFL makes the denoising performances of the model decrease. In terms of PSNR, the selected UNet from the denoising section reaches  26.917 versus 24.039 for the finetuned UNet with best PSNR. However, the initial goal of the DRFL is to improve the performance of the AFP system, and no longer to produce a clear denoised signal per se. This is why these results can be considered as promising for the specific context of audio fingerprinting.

    In the next section, we study the performances of the selected denoising model from \ref{Denoise}, and the best two models from this section on the test set.

    \chapter{Final results and discussion}

    \section{Final models}

    In the previous chapter, we selected three models that constitute potential great candidates to be integrated in the AFP system. In the following table \ref{fig:Final_table}, we synthesise their  performances on the validation set:  
  
    \begin{table}[H]
    \centering
    \includegraphics[width=1.0\textwidth]{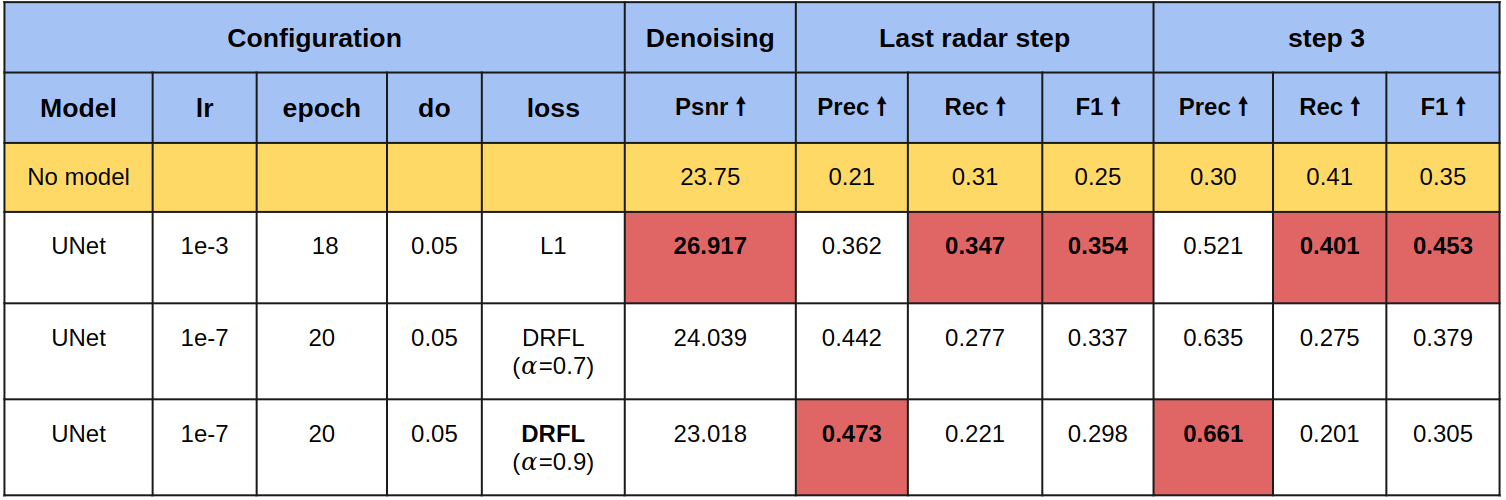}
    \noskipcaption{Final Models and Results on Validation Set}
    \label{fig:Final_table}
    \end{table}

    The first model comes from the group of experiments \textit{Learning to denoise}: it corresponds to the best denoising UNet, trained with \textit{L1 loss}, learning rate \textit{lr=1e-3} and \textit{dropout=0.05}. 
    
    As we can see in \ref{fig:Final_table}, the denoising UNet beats the "no model" baseline in all metrics on the validation set. In particular, it denoises the augmented CQTs,  increasing the PSNR of the denoised-clean CQTs pairs to 26.917 while it only reaches 23.75 for the clean-noisy pairs. When used in conjunction with Deezer's AFP system, the precision of the system in terms of preserved peaks increases to \textbf{0.521}, while the AFP system alone reaches \textbf{0.30}. The precision of the system is thus \textbf{multiplied by 1.7}, while the recall remains almost constant (0.41 vs 0.401 in the third RADAR step). This means that the percentage of extracted peaks from the noisy recordings by the AFP system belonging to the original music recording is multiplied by $1.7$.  

    The two other selected models were fine-tuned with the DRFL in order to \textit{learn to preserve peaks}. This allows to further increase the precision of the AFP system, reaching \textbf{0.635} for the first model and \textbf{0.661} for the second on the third RADAR step. In comparison to the "no model" baseline, the precision has been \textbf{multiplied respectively by 2.16 and 2.20}. 

    However, as explained in the previous chapter, this last improvement in precision is accompanied by a decrease in recall. Indeed, the more the system's precision increases during fine-tuning, the lower the recall gets. This forced us to select the final candidates using another criterion than only the precision. Here, we consider that good potential candidates correspond to models with a high precision and an F1-score (on the last RADAR step) over the "no model" baseline's: this means that the system's behaviour is still globally better than the baseline, while benefiting from a much higher precision. 

    More investigations should be done to study the exact effect of precision-recall trade off on the AFP system identification rate, but this is left for future works. 

    \section{Performances on test set}\label{test}

    We study the performances of the three selected candidates on the test set. The idea is to study how these models behave when confronted to other noise distributions than the one they were trained on. 

    For this, we define seven different augmentation pipelines used to test the model's robustness to various noise distributions. We want to see if the model can denoise the CQTs and improve the system's performances even when it is confronted to only one or some of the transformations of the augmentation pipeline it was trained on. 
    
    With the first three pipelines (\ref{fig:Pipeline_bn_light}, \ref{fig:Pipeline_bn_medium} and \ref{fig:Pipeline_bn_hard}), we only test the model robustness to \textbf{\gls{BN}}. These augmentation pipelines add additive noise to raw audios, using different \textit{Signal to Noise Ratios}. On the test set, the additive noise used to augment audios comes from the \textit{2020 evaluation DCASE dataset} and is thus different from the one used in the training pipeline (see \ref{BN} for more details). In the training pipeline, the models are confronted to hard constraints ($SNR \in [-10,-5]$). Here, we want to see if the models generalizes well when confronted to lighter constraints.
    
    \begin{table}[H]
    \centering
    \begin{minipage}{0.32\textwidth}
    \centering
    \captionsetup{justification=centering}
    \includegraphics[width=1\textwidth]{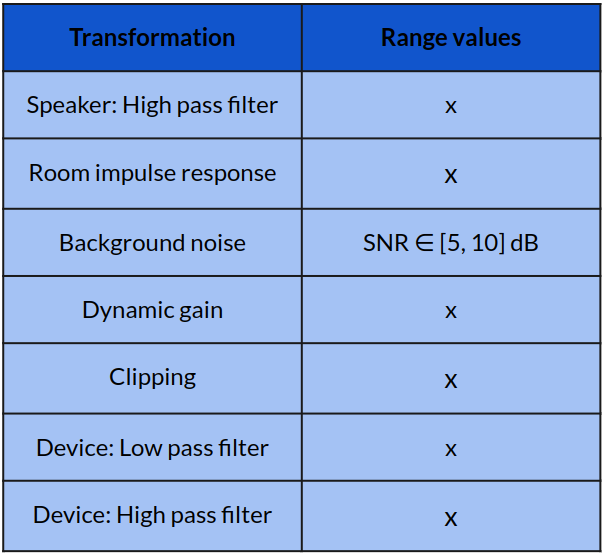}
    \caption{Light BN}
    \label{fig:Pipeline_bn_light}
    \end{minipage}\hfill
    \begin{minipage}{0.32\textwidth}
        \centering
        \captionsetup{justification=centering}
        \includegraphics[width=1\textwidth]
        {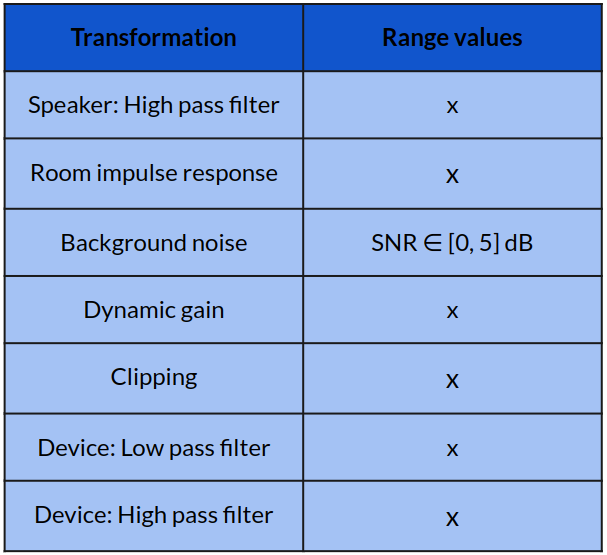}
        \caption{Medium BN}
        \label{fig:Pipeline_bn_medium}
    \end{minipage}
 \begin{minipage}{0.32\textwidth}
        \centering
        \captionsetup{justification=centering}
        \includegraphics[width=1\textwidth]
        {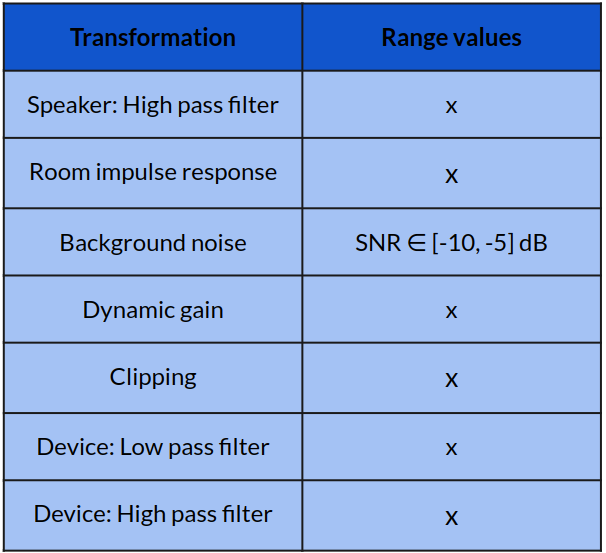}
        \caption{Hard BN}
        \label{fig:Pipeline_bn_hard}
    \end{minipage}
    \end{table}    

    We also test the model's robustness to reverberation. We define two pipelines: the first one adds only reverberation to the augmented samples \ref{fig:pipelineReverb}, while the second combines reverberation and strong background noise \ref{fig:Pipeline_reverb_bn}: 
    
    \begin{table}[H]
    \centering
    \begin{minipage}{0.5\textwidth}
    \centering
    \captionsetup{justification=centering}
    \includegraphics[width=0.7\textwidth]{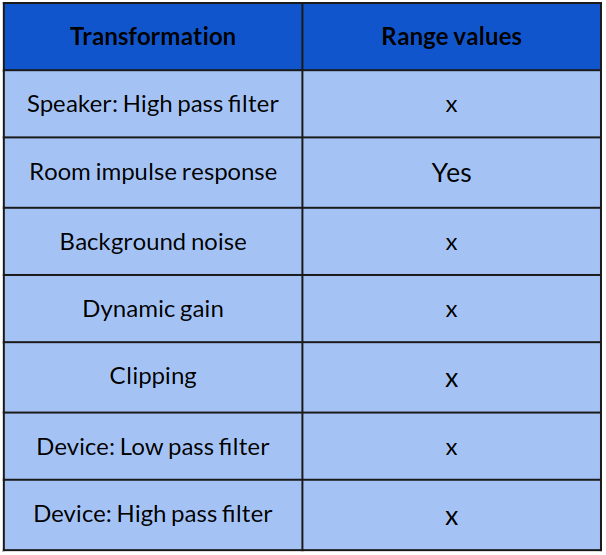}
    \caption{Reverberation}
    \label{fig:pipelineReverb}
    \end{minipage}\hfill
    \begin{minipage}{0.5\textwidth}
        \centering
        \captionsetup{justification=centering}
        \includegraphics[width=0.7\textwidth]
        {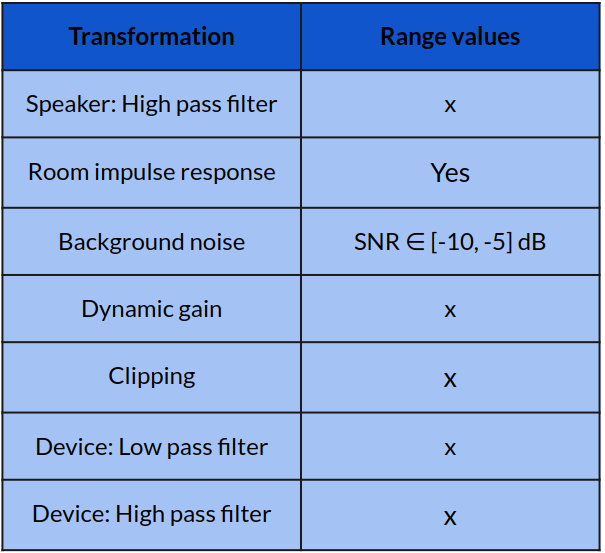}
        \caption{Reverberation + BN}
        \label{fig:Pipeline_reverb_bn}
    \end{minipage}
    \end{table}   

     The last two pipelines apply all the transformations used during training. We consider a "\textit{light}" and a "\textit{hard}" pipeline, in order to study the models behaviour to different levels of noise: 
     
    \begin{table}[H]
    \centering
    \begin{minipage}{0.5\textwidth}
    \centering
    \captionsetup{justification=centering}
    \includegraphics[width=0.7\textwidth]{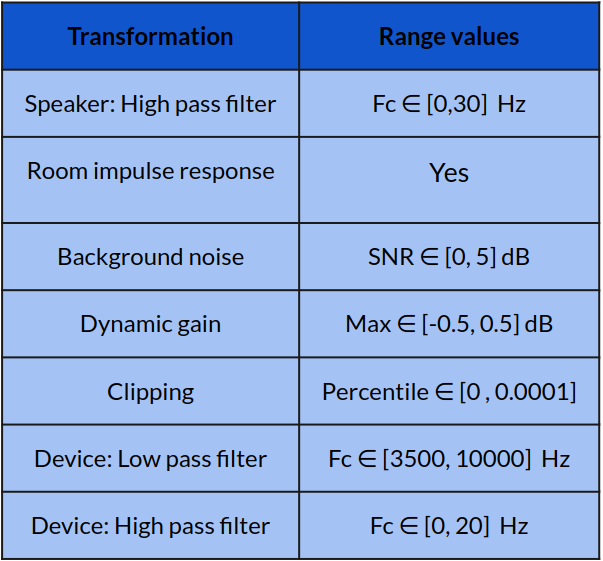}
    \caption{Full pipeline "Light"}
    \label{fig:LightConstraints}
    \end{minipage}\hfill
    \begin{minipage}{0.5\textwidth}
        \centering
        \captionsetup{justification=centering}
        \includegraphics[width=0.7\textwidth]{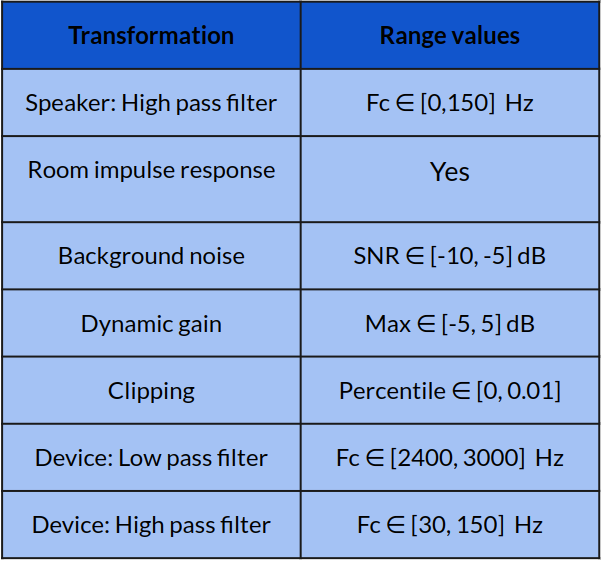}
        \caption{Full pipeline "Hard"}
        \label{fig:HardConstraints}
    \end{minipage}
    \end{table}   

    The results are presented in \ref{fig:PSNR_testset}, \ref{fig:Prec_testset} and \ref{fig:Rec_final_tab}. We rely on three metrics to assess the models performances: the first one corresponds to the \textbf{PSNR} computed on the denoised-clean CQTs pairs, and evaluates the models abilities to denoise CQTs. The second  and third metrics correspond to the \textbf{Precision} and \textbf{Recall} of the AFP system when used in conjunction with different models in terms of extracted peaks, computed on the third RADAR step. After each table, we briefly discuss the results.

    \clearpage
    \begin{itemize}
        \item \textbf{PSNR}:  
    \end{itemize}

    \begin{table}[H]
    \centering
    \captionsetup{justification=centering}
    \includegraphics[width=1.0\textwidth]{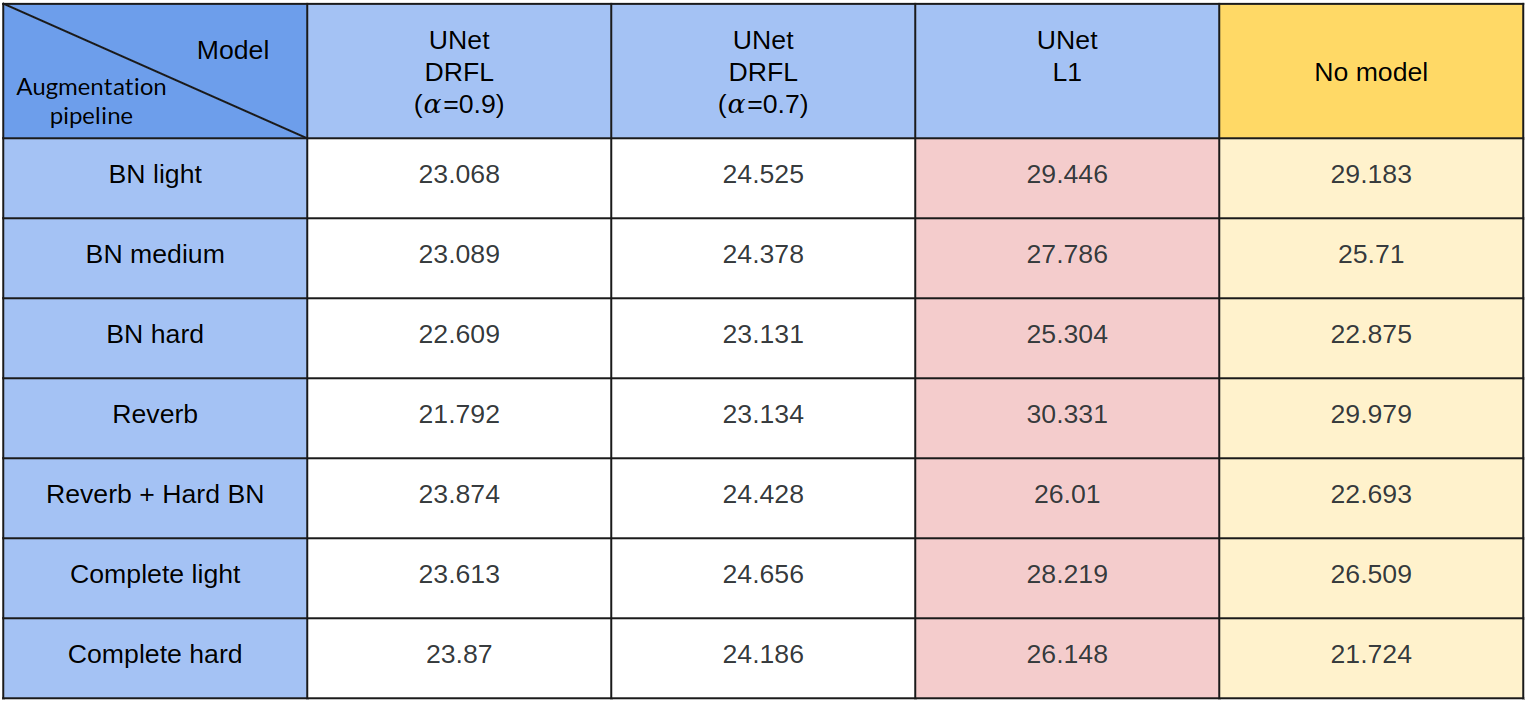}
    \noskipcaption{Influence of the Different Augmentation Pipelines on the PSNR, Results on Test Set}
    \label{fig:PSNR_testset}
    \end{table}

    From \ref{fig:PSNR_testset}, we can make several observations. First, the PSNR of the noisy-clean pairs, given by the "no model" column, depends on the strength of the applied transformations: the harder  the constraints are, the lower the PSNR is. Indeed, the \textit{complete hard} pipeline induces a PSNR of 21.724, whereas it reaches 26.509 for the light one. More generally, it seems that the background noise transformation is the one which impacts the PSNR more. Indeed, it reduces the PSNR from to 29.183 in \textit{BN light} to 22.875 in \textit{BN hard}, while the PSNR on the \textit{complete hard} pipeline is only one point lower (21.724). In comparison, adding reverb doesn't impact much the PSNR (29.979). 

    The Unet trained with L1 loss, or \textit{denoising UNet}, improves the PSNR  over the "no model" baseline in all the experiments. This suggests that even if trained on a specific noise distribution with hard constraints (corresponding to \textit{complete hard} pipeline), this model generalizes well enough to still denoise CQTs affected by other noise distributions, and more specifically less constraining distributions. 

    However, it seems that the denoising model can better denoise CQTs from the augmentation pipeline it was trained on. Indeed, it increases the PSNR from 21.724 to 26.148 on the \textit{complete hard} pipeline, while denoised CQTs from the \textbf{BN hard} pipeline reaches a PSNR of 25.304. Denoising CQTs augmented with \textit{BN hard} should be easier than the ones augmented with \textit{Complete hard}, as the latter are affected by 7 different transformations whereas the former undergo only one type of transformation. Yet, the model is able to better reconstruct CQTs augmented with the full pipeline.

    The two other models, fine-tuned using the DRFL do not behave well regarding denoising. Indeed, the PSNR of the predicted-noisy pairs is worst or close to the "no model" baseline for most augmentation pipelines. \textit{Complete hard} is the only pipeline for which there is a slight improvement over the "no model" baseline in terms of PSNR, increasing from \textbf{21.724} to \textbf{23.87} ( model trained with $\alpha=0.9$ ) and \textbf{24.186} ( model trained with $\alpha=0.7$). This is certainly due to the fact that the models were fine-tuned on samples augmented with this pipeline, starting from a denoising model and learning slowly to preserve peaks. 
    
    However, this small improvement is still quite below the PSNR reached by the denoising UNet (\textbf{26.148}). This confirms what we already observed on the validation set: the models fine-tuned with the DRFL loose their ability to denoise the CQTs, as they focus on peak preservation. 

    From these observations, we can see that even though the models can generalize to other noise distributions, they are less performant on noise distributions on which they weren't trained. This suggests to consider new training strategies regarding how audios are augmented during training. It would be for example interesting to apply only a single transformation on some samples of the train set rather than the full pipeline, to increase the model's robustness to specific transformations. This could in the end also help increase the model's robustness to the full pipeline. Investigating such training strategies is left for future works. 
    
    \begin{itemize}
        \item \textbf{Precision and Recall}:  
    \end{itemize}

    \begin{table}[H]
    \centering
    \captionsetup{justification=centering}

    \includegraphics[width=1.0\textwidth]{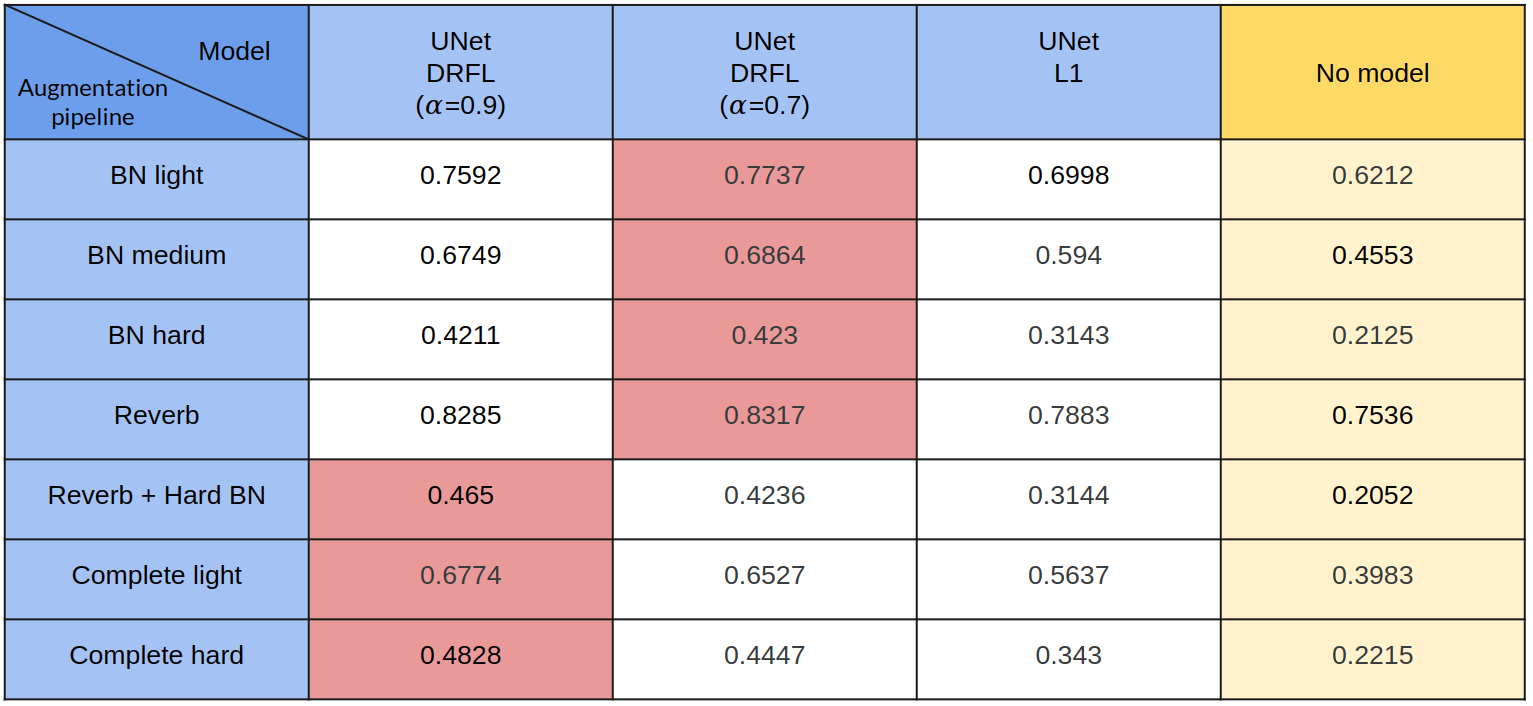}
    \noskipcaption{Influence of the Different Augmentation Pipelines on the AFP system Precision (RADAR third step), Results on Test Set}
    \label{fig:Prec_testset}
    \end{table}

    \begin{table}[H]
    \centering
    \captionsetup{justification=centering}

    \includegraphics[width=1.0\textwidth]{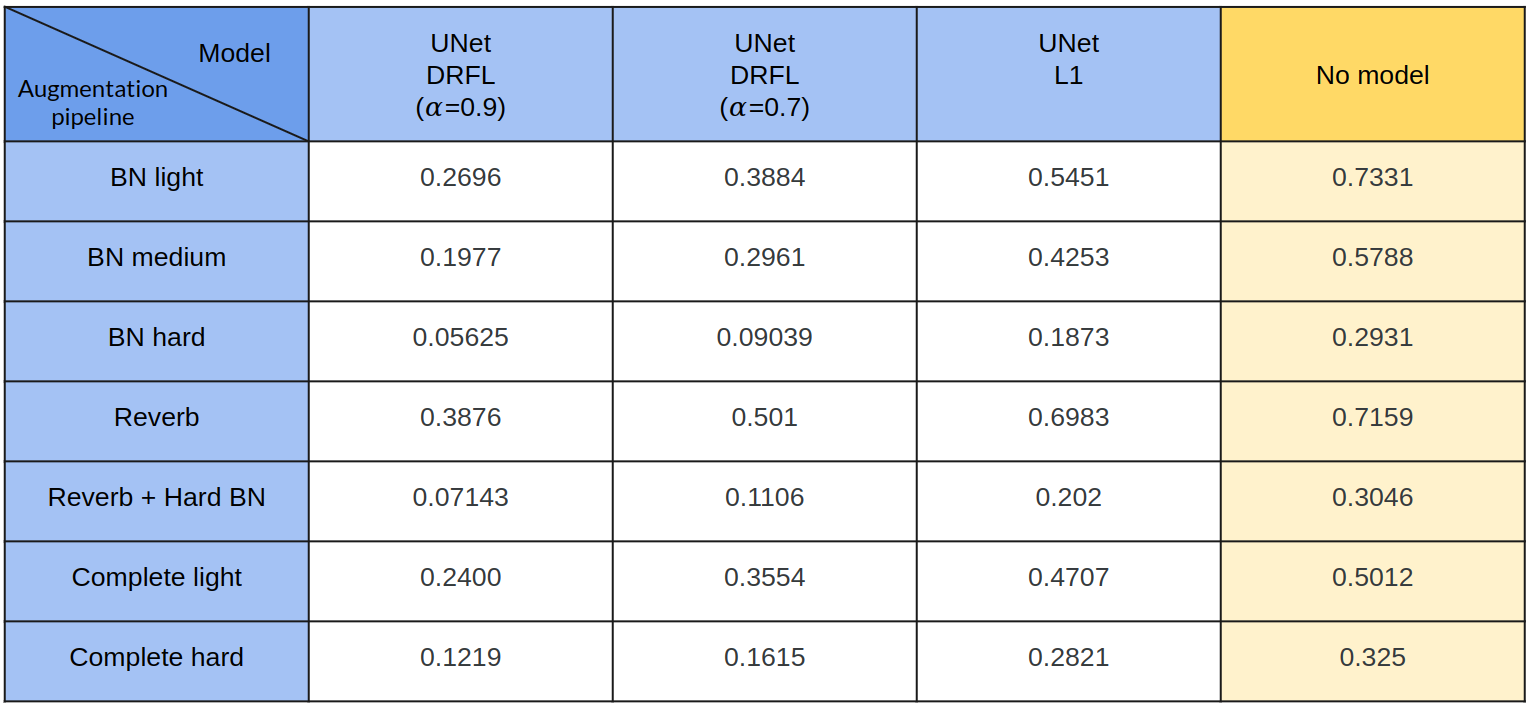}
    \noskipcaption{Influence of the Different Augmentation Pipelines on the AFP system Recall (RADAR third step), Results on Test Set}
    \label{fig:Rec_final_tab}
    \end{table}

    We rely on the third RADAR step to evaluate the precision and recall of the AFP system when used in conjunction with the DL models. We compute these metrics on this intermediate RADAR step rather than the last one as it can better highlight the models behaviour, with significant differences between good and bad behaviours. We can make several observations:

     First, the three models \textbf{have a better precision than the "no model" baseline for all augmentation pipelines}. This confirms their ability to preserve spectral peaks and thus be beneficial to the AFP system. In particular, we increase the precision of the system from 0.2215 to 0.4828 on the \textit{Complete hard} and from 0.39 to 0.67 on the \textit{Complete light} pipeline. 
     
     However, it seems that \textbf{recall decreases over the "no model" baseline for all models}. This was expected for the UNet fine-tuned with the DRFL but not for the one trained to denoise with L1 loss. The recall of the denoising UNet was indeed similar to the one of the "no model" baseline on the validation set. This suggests that this model do not generalize as well as it should on new noise distributions. This may be due to a lack of variability in the augmentation pipeline during training. Nevertheless, the results obtained on the \textit{complete light} and \textit{complete hard} pipelines are not so shocking. The recall of the denoising UNet is still pretty close to the one of the no model baseline, even if it is a bit below: On the \textit{Complete hard}, the recall decreases from 0.325 to 0.2821, while it goes from 0.5012 to 0.4707 on the \textit{Complete light} pipeline. This, in addition to the important increase in terms of precision of the denoising UNet over the baseline (0.2215 to 0.343 on the \textit{Complete hard} pipeline for example), suggests that \textbf{the AFP could clearly benefit from the denoising model}.

    Finetuning models with \textbf{the DRFL further increases precision but decreases recall}. In particular, when audios are augmented using the complete data augmentation pipelines, we have: $\textit{Precision}_{\alpha=0.9} > \textit{Precision}_{\alpha=0.7} > \textit{Precision}_{L1}$ and $Rec_{\alpha=0.9} < Rec_{\alpha=0.7} < Rec_{L1}$ which is coherent with what we observed on the validation set. The different models behave as they are supposed to when confronted to new audios segments from noise distributions including all the transformations. As the augmentation pipelines \textit{Complete light} and \textit{Complete hard} are the most representative of realistic noises present in real environments, this suggests that the different models could generalize sufficiently to be integrated in AFP systems used in typical \textit{SongCatcher} scenarios.

    As we already observed in \ref{fig:PSNR_testset}, the models doesn't behave as well as they should on CQTs augmented with only one or some transformations. This is due to the fact that they were not trained to do so. We note an important decrease in recall for all models. When confronted to \textit{BN hard}, for example, the recall decreases from 0.29 for the 'no model' baseline to 0.18 with the denoising model. It then reaches values that are much too low with the DRFL models, with recalls under 0.10. In terms of precision, we can see that even though the model trained with the DRFL and $\alpha=0.9$ has a better precision than the one trained with $\alpha=0.7$ on the validation set, it is no longer the case when single transformations are applied to the audios. Indeed, in this case, the UNet fine-tuned with parameter $\alpha=0.7$ gets the best precision (altough the precision of both models is pretty close). Again, this suggests to consider training pipelines with audios augmented using only one or some transformations of the augmentation pipeline rather than all of them at the same time.

    \clearpage
    \section{Future works}

    Altough our deep learning models show encouraging results in terms of capacity to denoise CQTs and ability to preserve spectral peaks, we still haven't study the impact of their integration in the \textbf{AFP system's identification rate}. By increasing the percentage of retrieved spectral peaks  belonging to clean music references, we should logically increase the system's identification rate. However, establishing the exact relation between these two metrics would require more investigations. In particular, the \textit{precision-recall} trade off might have an important impact on the system. This corresponds to our next direction of work, which is necessary to get a more concrete idea of how much the AFP system could be improved with our deep learning models.  
    
    In the longer term, we would like to improve our current methodology to build better models, that can better denoise and preserve spectral peaks while generalizing to more music recordings with different noise distributions. Several research directions are considered:
    
    A first way to improve the DL models concerns \textbf{the data augmentation pipeline}, and more precisely how we augment audios during training. In this project, we only consider single clean-augmented pairs for each music track segment: the augmented CQTs are computed before training, stored, and fed to the model at training time.  Another way to train the models could be to augment the clean audios directly during training, so that models never see twice the same noise associated to the same music extract. This would generate more variability in the noise handled by the model. Each audio segment would be then augmented in a different way at each epoch, making the model better understand how the same music can be affected differently. The main challenge of this new training strategy would be to make the augmentation pipeline sufficiently fast in order to use it online. 

    As we noticed in \ref{test}, the current DL models can well denoise and preserve spectral peaks of audios affected by the complete audio augmentation pipeline, but they can struggle when confronted to single transformations. This can be problematic for production level applications, as some sources of noises considered in our augmentation pipeline might be missing or limited in a real environment. To solve this problem, one approach could be to augment some audios of the training set using only one or some of the transformations of the augmentation pipeline rather than all of them at once.   

    In order to continue improving the data augmentation pipeline, we could also work on refining its modelling. Indeed, in this project, we modeled some sources of noise in the augmentation pipeline in a rather simplistic way. Speakers, for example, are represented using first order high pass filters. Modeling them in a more complex way could help better simulate the noise we want to remove from recordings, and thus build better denoising models. Also, we could use larger background noise and Room Impulse Responses datasets, which would increase the variability of the noisy samples processed by the model.

    Another research direction to improve our models would be to consider \textbf{other training strategies}. In particular, our models could be trained with adversarial losses. As we saw in \ref{losses}, GANs are often used in the SE literature. Encoder-Decoders trained with GANs have indeed state of the art performances today in SE, and have shown a clear superiority over models trained with simple $\textit{L}_{p}$ losses. The specificity of these models is they are trained to make the denoised outputs look realistic. As we can see in \ref{fig:CAD_2} and \ref{fig:CAD_3}, even if our current models can reconstruct music notes in their predicted CQTs, there is still a clear difference between the original clean spectrograms and the reconstructed one in terms of visual appearance: the predicted CQTs have more dark areas. Training with GANs could thus help reduce these visual artefacts, improving potentially both denoising capacity and system's ability to preserve spectral peaks. 
    
    Finally, one other research direction could be to improve the current DRFL. This loss is currently not perfectly optimised in terms of engineering, which limits its use during training to the first two RADAR steps only. We could work on making the computations faster and further study how to optimally use the FTL during training.

    \chapter{Conclusion}

    In this project, we introduce a new hybrid strategy to help a peak-based AFP system improve its robustness to background noise recorded in real environments. We incorporate a deep learning denoising model in the AFP system's pipeline, placing it in front of a spectral peak extraction algorithm.

    To train our model, we define a realistic augmentation pipeline simulating typical noises found in places where applications such as \textit{SongCatcher} are used, and augment a database consisting of more than 48K music audios extracts. This allows us to train multiple EncoderDecoder architectures on a dataset of clean-noisy CQTs pairs. 
    
    With the best model trained to denoise CQTs, a UNet, we are able to increase the PSNR on the validation set from 23.75 to 26.917. This, visually, corresponds to predicted CQTs that look quite similar to their clean references, with the notable presence of reconstructed notes. This allows to multiply the precision of the AFP system in terms of preserved spectral peaks by 1.5. 

    We then investigate a training strategy based on a new loss function designed to help the DL models learn to preserve spectral peaks. This further multiplies the system's precision by 2 over the "no model" baseline on the validation set. However, this increase in precision is accompanied by a decrease in recall. Moreover, experiments on the test set have shown that our models may have trouble generalizing to new noise distributions unseen during training. 

    Future works will seek to study the impact of the \textit{precision-recall} trade off on the AFP system's identification rate, as well as explore new training strategies to improve the model's generalization capacity. 

    \clearpage
    \appendix
    \chapter{Appendix}

    \section*{4 - Methodology}

    \subsubsection*{4.2 - Constructing a dataset with great musical variety}

    \begin{figure}[H]
    \centering
    \includegraphics[width=0.8\textwidth]{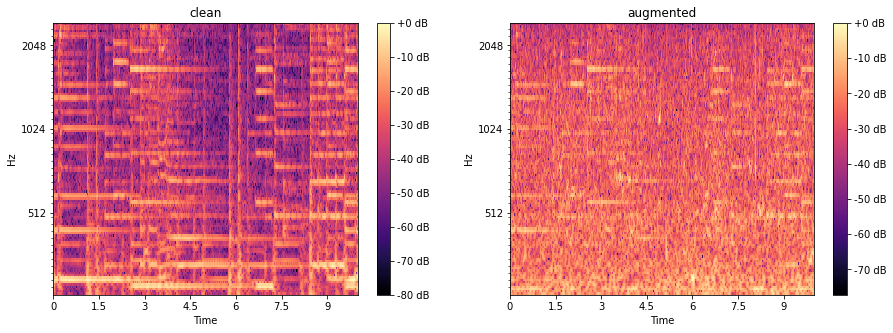}
    \noskipcaption{A Clean CQT and its Augmented Version }
    \label{fig:Clean-NoisyPairs2}
    \end{figure}

    \subsection*{4.3 - Selecting potential Encoder-Decoder architectures}

    \begin{figure}[H]
    \centering
    \includegraphics[width=0.8\textwidth]{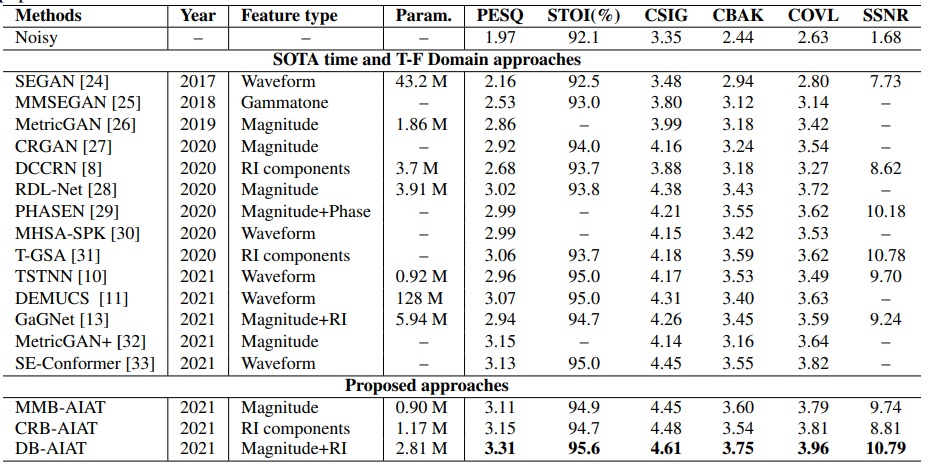}
    \noskipcaption{Comparison of AIAT to Other Models}
    \label{fig:SOTA_AIAT}
    \end{figure}
    
    \clearpage

    \section*{5 - Trainings}

    \subsection*{5.2 - Visualising the denoised CQTs}
    
    \begin{figure}[H]
    \centering
    \includegraphics[width=0.75\textwidth]{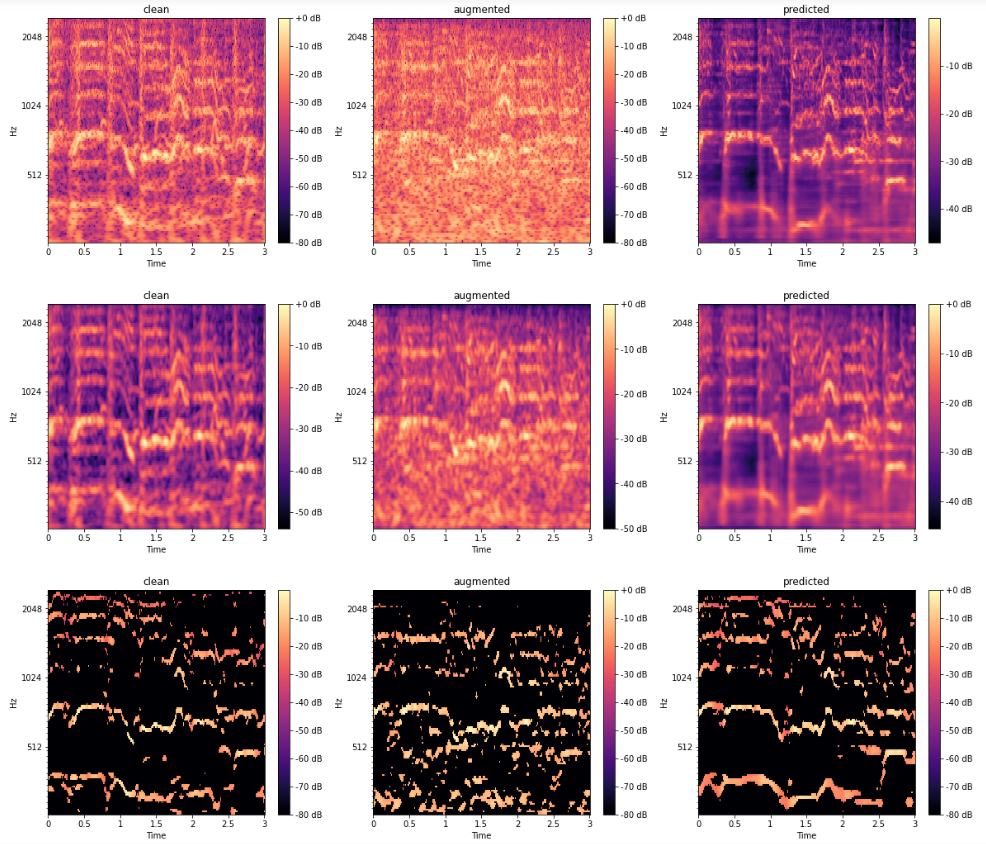}
    \noskipcaption{Clean, Augmented and Denoised CQTs }
    \label{fig:CAD_2}
    \end{figure}

    \begin{figure}[H]
    \centering
    \includegraphics[width=0.75\textwidth]{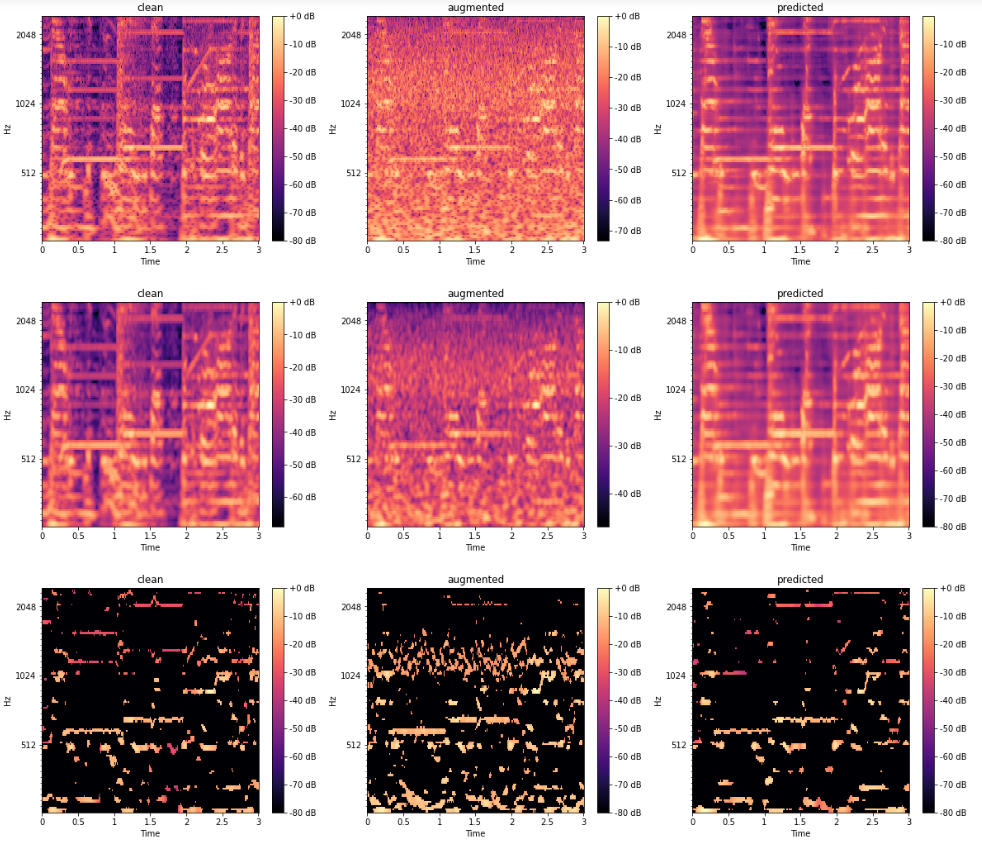}
    \noskipcaption{Clean, Augmented and Denoised CQTs}
    \label{fig:CAD_3}
    \end{figure}

    \subsection*{5.3 - Learning to preserve peaks}

    \begin{figure}[H]
    \centering
    \includegraphics[width=1.0\textwidth]{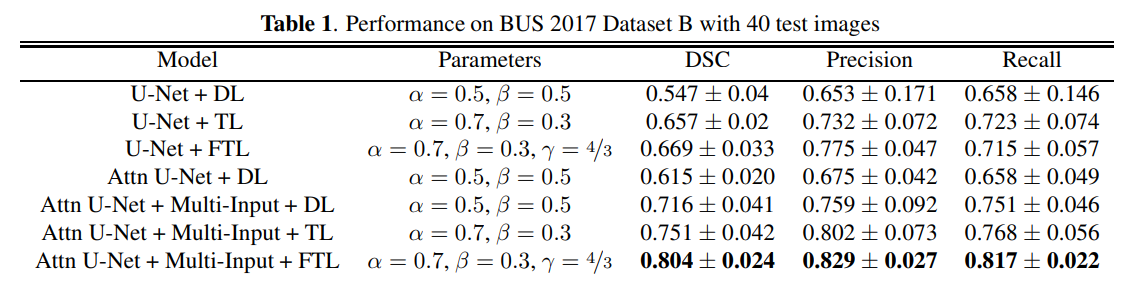}
    \noskipcaption{Influence of the FTL Hyperparameters, results on BUS2017 dataset \cite{FocalTverskyLoss}}
    \label{fig:FTL_table}
    \end{figure}

    \clearpage
    \printbibliography
    
    \clearpage

\end{document}